\pdfoutput=1
\documentclass[aps,prl,twocolumn,superscriptaddress,nofootinbib]{revtex4-2}
\usepackage{amsmath,amssymb}
\usepackage{graphicx}
\usepackage[hidelinks]{hyperref}
\newcommand{\Tr}{\mathrm{Tr}}
\newcommand{\E}{\mathbb{E}}

\begin{document}

\title{Efficiency-Induced Freezing in Quantum-State Purification}

\author{Jiaxin Liu}
\affiliation{School of Instrumentation and Optoelectronic Engineering, Beihang University, Haidian, Beijing 100191, China}
\affiliation{Institute of Large-Scale Scientific Facility, Beihang University, Beijing 100191, China}
\affiliation{Hangzhou Institute of Extremely-Weak Magnetic Field Major National Science and Technology Infrastructure, Hangzhou 310051, China}

\author{Zuoxian Wang}
\affiliation{School of Instrumentation and Optoelectronic Engineering, Beihang University, Haidian, Beijing 100191, China}
\affiliation{Institute of Large-Scale Scientific Facility, Beihang University, Beijing 100191, China}

\author{Feng Li}
\affiliation{Institute of Large-Scale Scientific Facility, Beihang University, Beijing 100191, China}
\affiliation{Hangzhou Innovation Institute, Beihang University, Hangzhou 310051, China}

\author{Danyue Ma}
\affiliation{School of Instrumentation and Optoelectronic Engineering, Beihang University, Haidian, Beijing 100191, China}
\affiliation{Institute of Large-Scale Scientific Facility, Beihang University, Beijing 100191, China}
\affiliation{Hangzhou Institute of Extremely-Weak Magnetic Field Major National Science and Technology Infrastructure, Hangzhou 310051, China}

\date{\today}

\begin{abstract}
Any nonzero detection loss qualitatively changes feedback-controlled
purification under diffusive monitoring.  In every finite dimension, we prove
a sharp, dimension-independent ceiling on the decay of trajectory-averaged
impurity moments, uniformly over admissible predictable feedback protocols.
Below unit efficiency, this ceiling becomes independent of moment order above a
critical value and is attained on extremal rank-two quantum-nondemolition (QND)
faces.  For generic observable spectra, a determinant-root law precludes every
full-rank state from attaining this boundary rate over an explicit moment-order
interval.  For qubits at $0<\eta<1$, the frozen rate is the exact optimum, set by
rare, persistently mixed trajectories.  Parameter-free finite-action scaling
functions resolve both the rounded QND moment-order transition and the
near-unit QND--always-unbiased crossover.
\end{abstract}

\maketitle

Continuous measurement drives a monitored state toward purity.  Hamiltonian feedback
can accelerate selected purification objectives~\cite{Jacobs2003,CombesJacobs2006,CWJ2010,
WisemanMilburn2010}, while measurement-based feedback also provides rigorous
state-reduction and stabilization guarantees~\cite{vanHandel2005,TicozziViola2008,
Cardona2020}.  Superconducting-qubit trajectories
have been reconstructed~\cite{Murch2013} and steered by real-time stroboscopic
feedback~\cite{Campagne2013}; purification dynamics also probes measurement-induced
transitions in monitored many-body systems~\cite{GullansHuse2020}.  Detection
efficiency is the decisive resource: observed backaction generates a record available
for feedback, whereas the unobserved component causes dephasing without information.

For a qubit at unit efficiency, an unbiased-basis feedback protocol minimizes the
finite-horizon mean impurity~\cite{Jacobs2003,WisemanBouten2008}, whereas a fixed QND
protocol minimizes the mean first-passage time to a purity
threshold~\cite{WisemanRalph2006,WisemanBouten2008,Li2013}.  R\'enyi objectives
distinguish local and global optima~\cite{TeoCombesWiseman2014}.  The impurity-moment
family considered here interpolates between low-order sensitivity to nearly pure
trajectories and high-order weight on rare mixed records.  Under detection
inefficiency, Markovian-feedback state preparation was studied in
Ref.~\cite{CombesWiseman2011}, and finite-horizon costs were optimized numerically in
Ref.~\cite{Jiang2020}.  Without extrinsic decoherence, Li \emph{et al.}\
proved QND globally optimal for first passage at every efficiency and for
finite-horizon mean impurity when $\eta\leq1/2$, leaving $1/2<\eta<1$
open~\cite{Li2013}.  Other bounds address fixed-instrument purification, estimator
contraction, eigenvalue-permutation feedback, and fixed-QND collapse
rates~\cite{Bompais2026,CombesWisemanJacobs2008,BenoistPellegrini2014}.  Yet no result gives
a protocol-uniform ceiling on trajectory-averaged impurity moments under admissible
predictable, noncommuting control.  Local-filter covariance and channel-factorization laws for
minimal-purification G-concurrence~\cite{Gour2005,Konrad2008,Tiersch2008,Gour2010},
together with monitored average-concurrence
studies~\cite{Viviescas2010,VogelsbergerSpehner2010,GuevaraViviescas2014}, motivate the
determinant law.  At unit efficiency, the averaged determinant root obeys a
protocol-independent law; detection inefficiency introduces a nonnegative
conditional correction.

\begin{figure*}[t]
\centering
\includegraphics[width=\textwidth]{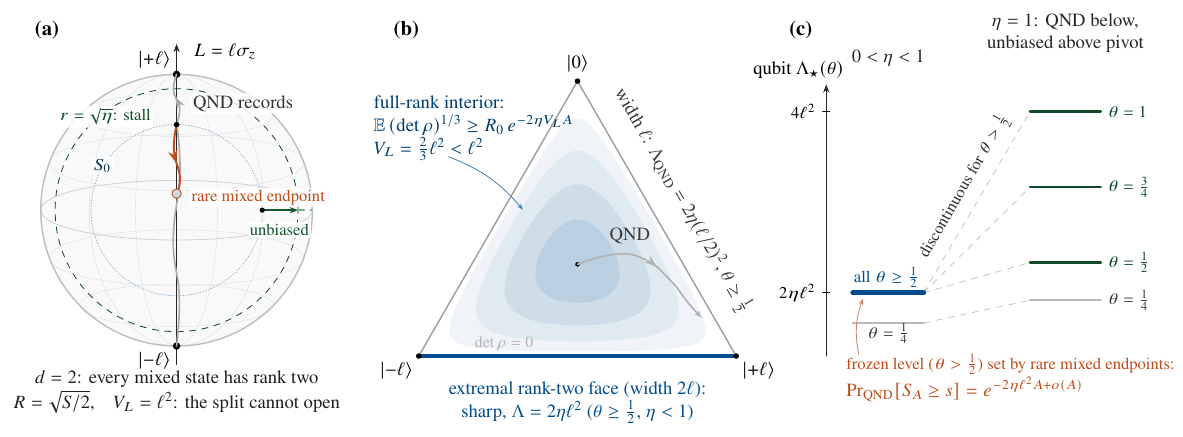}
\caption{Geometry of efficiency-induced freezing (trajectories are schematic;
determinant contours and exponents are exact).
(a) For a qubit, QND records diffuse toward the poles, an unbiased protocol stalls
at $r=\sqrt\eta$, and a rare mixed endpoint remains near the center.
(b) On the commuting population slice of a qutrit, the extremal rank-two QND face
attains the ceiling for $\theta\geq\tfrac12$ and $0<\eta<1$.  The determinant-root
law, $R_0=(\det\rho_0)^{1/3}$, precludes this rate in the full-rank interior for
$\tfrac12\leq\theta<1$; a QND path reaches the boundary only asymptotically.
(c) Level scheme for the optimal qubit exponent.  For $0<\eta<1$, every order
$\theta\geq\tfrac12$ freezes at $2\eta\ell^2$.  Rare mixed endpoints set this level:
for a QND channel aligned with any mixed initial state ($\ell>0$),
$\Pr_{\rm QND}[S_A\geq s]=e^{-2\eta\ell^2A+o(A)}$ for every fixed threshold
$s\in(0,\tfrac12)$.  At unit efficiency, unbiased feedback instead gives
$4\theta\ell^2$ above the pivot, discontinuously as $\eta\to1^-$.}
\label{fig:mechanism}
\end{figure*}

This Letter establishes a sharp, dimension-independent, action-based purification
speed limit for impurity moments---an analogue, in spirit, of a quantum speed
limit~\cite{DeffnerCampbell2017,GarciaPintos2022,delCampo2013,Taddei2013}.  For every
inefficient detector, the protocol-uniform asymptotic ceiling freezes above a
critical moment order and is attained on rank-two QND boundary faces.  A
determinant-root law yields a strictly lower ceiling for generic full-rank states
over an explicit moment-order interval.  For a qubit, the protocol-uniform bound is
the exact optimum: rare, persistently mixed trajectories set its frozen branch,
producing the boundary--interior dichotomy in Fig.~\ref{fig:mechanism}.

\emph{Controlled measurement.---}
Let a finite-dimensional conditional state obey the It\^o stochastic master
equation~\cite{Belavkin1992,GisinPercival1992,BarchielliGregoratti2009,JacobsSteck2006}
\begin{equation}
d\rho=-i[H_t,\rho]dt+M\mathcal D[L]\rho\,dt
 +\sqrt{\eta M}\,\mathcal H[L]\rho\,dW,
\label{eq:sme}
\end{equation}
with $L=L^\dagger$, $\mathcal D[L]\rho=L\rho L-\tfrac12\{L^2,\rho\}$,
$\mathcal H[L]\rho=L\rho+\rho L-2\Tr(L\rho)\rho$, and half-spread
$\ell=(\lambda_{\max}-\lambda_{\min})/2$; adding a multiple of $\openone$ to $L$
does not affect Eq.~\eqref{eq:sme}.  The impurity is $S=1-\Tr\rho^2$.  The
orientation of $L$ and the finite-variation Hamiltonian $H_t$ may both be
predictable functions of the record.  We use the accumulated action
$A=\int_0^tM(s)ds$, assumed to grow without bound, as a deterministic clock
($dW_A^2=dA$)~\cite{RevuzYor1999}.  We evaluate $S_A$ at the inverse-action time
$\tau_A=\inf\{t:\int_0^tM(s)ds\geq A\}$; under this assumption, a scheduled strength
only reparametrizes time.  Hamiltonian control preserves the spectrum of $\rho$
pathwise and affects $S$ only through the next measurement orientation.  The control
class allows arbitrarily fast Hamiltonian basis rotations but no additional
dissipative control channel.
Simultaneous complementary channels~\cite{Ruskov2012} and
measurement-current-proportional feedback~\cite{WisemanMilburn1993fb} define
different resource models.

For a qubit, $S=2\det\rho\in[0,\tfrac12]$.  Both its drift and squared diffusion are
affine in the orientation parameter $x=\tfrac14(L_{11}-L_{22})^2\in[0,\ell^2]$,
where $x=\ell^2$ is aligned (QND) and $x=0$ is unbiased [Supplemental Material
(SM)~\cite{SM}].  Endpoint generator inequalities therefore apply to every
predictable orientation.  The endpoints represent persistent readout and contraction
to a mixed fixed point, respectively; the asymptotic analysis determines the
long-action optimum.

\begin{figure*}[t]
\centering
\includegraphics[width=\textwidth]{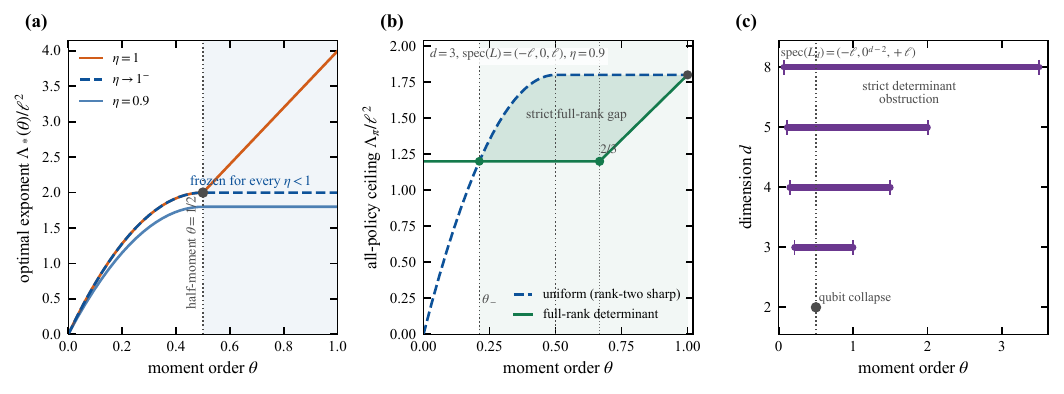}
\caption{Boundary freezing and the full-rank obstruction.
(a) Exact qubit optimum.  For $0<\eta<1$, it freezes at $2\eta\ell^2$ above
$\theta=\tfrac12$ (blue); at unit efficiency, unbiased feedback gives
$4\theta\ell^2$ (orange).  The dashed curve is the nonuniform limit
$\eta\to1^-$.  (b) For the canonical qutrit spectrum $(-\ell,0,\ell)$ at
$\eta=0.9$, rank-two QND faces attain the ceiling (dashed blue).  The
determinant-root law gives the stronger full-rank bound (green) for
$\theta_-<\theta<1$, where $\theta_-=(1-1/\sqrt3)/2$.  (c) For $0<\eta<1$ and
$L_d=\operatorname{diag}(-\ell,0^{d-2},+\ell)$, the strict full-rank obstruction
occupies $\theta_-^{(d)}<\theta<\theta_+^{(d)}$, where
$\theta_-^{(d)}=[1-\sqrt{1-2/d}]/2$ and $\theta_+^{(d)}=(d-1)/2$.  The interval
collapses to the half-moment pivot at $d=2$ and opens for every $d>2$.}
\label{fig:freezing}
\end{figure*}

For a protocol $\pi$ and moment order $\theta>0$, define the moment exponent and its
optimum
\begin{equation}
\begin{aligned}
\Lambda_\pi(\theta,\eta)&=\liminf_{A\to\infty}
-\frac1A\ln\E_\pi[S_A^\theta],\\
\Lambda_\star(\theta,\eta)&=\sup_\pi\Lambda_\pi(\theta,\eta),
\end{aligned}
\label{eq:exponent}
\end{equation}
suppressing the $(d,\rho_0)$ dependence.  In every finite dimension, for every mixed
initial state and every admissible predictable feedback protocol at $0<\eta<1$, we have
\begin{equation}
\Lambda_\pi(\theta,\eta)\leq\Lambda_{\rm univ}(\theta,\eta):=
\begin{cases}
8\theta(1-\theta)\eta\ell^2,&0<\theta\leq\tfrac12,\\[2pt]
2\eta\ell^2,&\theta\geq\tfrac12,
\end{cases}
\label{eq:alldceiling}
\end{equation}
and both constants are sharp uniformly.  An embedded rank-two QND face attains both
branches, but not every full-rank state with $d>2$ can do so.  For a fixed unitary
orbit of $L$, define the spectral variance
\begin{equation}
V_L:=\frac1d\Tr\!\left(L-\frac{\Tr L}{d}\openone\right)^2,
\qquad \alpha_d:=\frac{d-1}{d}.
\label{eq:spectralvariance}
\end{equation}
Every full-rank state obeys the determinant bound
\begin{equation}
\Lambda_\pi(\theta,\eta)\leq
2\eta V_L\max\!\left\{1,\frac{\theta}{\alpha_d}\right\},
\label{eq:detceiling}
\end{equation}
which supersedes Eq.~\eqref{eq:alldceiling} when smaller.  Generically
$V_L<\ell^2$; equality requires a balanced dichotomic spectrum, with equal endpoint
multiplicities and even $d$.  Thus, under a generic observable, every full-rank
$d>2$ state remains strictly below the rank-two constant at the half-moment pivot
$\theta=\tfrac12$.  For a mixed qubit, the matching law holds from every initial
state and gives the exact optimum
\begin{equation}
\Lambda_\star(\theta,\eta)=
\begin{cases}
8\theta(1-\theta)\eta\ell^2,
 &0<\theta<\tfrac12,\quad 0<\eta\leq1,\\[2pt]
2\eta\ell^2,
 &\theta\geq\tfrac12,\quad 0<\eta<1,\\[2pt]
4\theta\ell^2,
 &\theta\geq\tfrac12,\quad \eta=1.
\end{cases}.
\label{eq:phase}
\end{equation}
The all-dimensional ceiling and the matching QND law give the frozen second branch
[Fig.~\ref{fig:freezing}(a)].  At unit efficiency, unbiased feedback instead attains
the high-order optimum.  No ergodicity is assumed.  At
$(\theta,\eta)=(1,1)$, the result recovers the
Jacobs--Wiseman--Bouten theorem.  For $\theta=1$ and
$\tfrac12<\eta<1$, it determines the globally optimal long-action exponent.  This
asymptotic result echoes---but does not resolve---the open finite-horizon problem of
Li \emph{et al.}

\emph{A dimension-independent half-moment limit.---}
Finite-action bounds determine the low-order branch and the special role of
$\theta=\tfrac12$.  Center the observable as
$\widetilde L=L-\Tr(L\rho)$ and introduce
\begin{equation*}
T_1=\Tr(\widetilde L^2\rho^2),\quad
T_2=\Tr(\widetilde L\rho\widetilde L\rho),\quad
T_3=\Tr(\rho^2\widetilde L).
\end{equation*}
For a qubit, the exact identities $T_3^2=x(1-2S)S^2$ and
$T_2S+T_3^2=\ell^2S^2$ hold for every measurement orientation.  Thus the quartic
relation is saturated throughout the qubit control interval, whereas only the
corresponding inequality extends to arbitrary dimension.  In the action clock,
Eq.~\eqref{eq:sme} gives
\begin{equation}
dS=2[(1-\eta)T_1-(1+\eta)T_2]dA-4\sqrt\eta\,T_3dW_A.
\label{eq:Sgeneral}
\end{equation}
The commutator contributes no spectral drift; the final $dW_A$ term is the
innovation from the observed record, and centering $L$ exposes the competition
between observed and unobserved backaction.

The SM proves the quartic trace inequality $T_2S+T_3^2\leq\ell^2S^2$ and the
pairing bound $T_3^2\leq\ell^2S^2$.  Combining these bounds without discarding the
efficiency-dependent $T_1$ term yields the master inequality
\begin{equation}
(1+\eta)T_2-(1-\eta)T_1+2\eta\frac{T_3^2}{S}
\leq2\eta\ell^2S.
\label{eq:master}
\end{equation}
These feedback-independent spectral statements have optimal uniform constants.
Apart from $S=0$ or $\ell=0$, the quartic inequality is saturated if and only if
$\operatorname{rank}\rho=2$ and the support compression $P_\rho L P_\rho$ has width
$2\ell$ [SM~\cite{SM}].  A diagonal two-level mixture of the extremal eigenvectors of
$L$ realizes this QND face.  At $\eta=1$, this is precisely the equality condition
for the pointwise half-moment ceiling: every mixed qubit saturates the bound for every
orientation, whereas the inequality is strict for every state of support rank at
least three.
This pointwise dichotomy becomes a uniform long-action exponent gap whenever
$V_L<\ell^2$, by the determinant law below.  The bounds are independent of dimension
and instantaneous eigenvalue multiplicities.

For every fixed $\theta>0$, It\^o's formula and Eq.~\eqref{eq:master} give, for the
action-clock generator $\mathcal A$,
\begin{equation}
-\frac{\mathcal A S^\theta}{S^\theta}
\leq4\theta\eta\ell^2
 +4\theta(1-2\theta)\eta\frac{T_3^2}{S^2}.
\label{eq:momentgenerator}
\end{equation}
For $\theta\leq\tfrac12$ the pairing bound controls the last term; for
$\theta\geq\tfrac12$ it can be discarded.  Gr\"onwall's inequality then gives the
protocol-independent, finite-action envelope
\begin{equation}
\begin{split}
\E[S_A^\theta]&\geq S_0^\theta
\exp[-c(\theta)\eta\ell^2A],\\
c(\theta)&=
\begin{cases}
8\theta(1-\theta),&\theta\leq\tfrac12,\\
4\theta,&\theta\geq\tfrac12.
\end{cases}.
\end{split}
\label{eq:envelope}
\end{equation}
This finite-action bound holds at every $A$.  QND asymptotically attains its
low-order exponent for every $0<\eta\le1$, with exact finite-action equality
at $\theta=\tfrac12$.  At unit efficiency, unbiased feedback attains the high-order
branch exactly.  At the pivot, the bound becomes
\begin{equation}
\E[\sqrt{S_A}]\geq\sqrt{S_0}\,e^{-2\eta\ell^2A}.
\label{eq:rootbound}
\end{equation}
On a rank-two QND face, Eq.~\eqref{eq:rootbound} is an equality for every $A$; the
same construction embeds the qubit QND spectrum in any finite dimension.  For
$0<\theta\leq\tfrac12$, it matches the envelope in exponent at every efficiency.
At unit efficiency, an embedded unbiased channel attains the high-order
branch $4\theta\ell^2$.  For $0<\eta<1$ and $\theta>\tfrac12$, the
change-of-measure theorem below sharpens the asymptotic ceiling from
$4\theta\eta\ell^2$ to $2\eta\ell^2$, attained on the rank-two QND face.

For $\theta<\tfrac12$ the positive It\^o correction in Eq.~\eqref{eq:momentgenerator}
is controlled by the maximal innovation $|T_3|$; for $\theta>\tfrac12$ its sign
reverses and fluctuations penalize the moment.  The cancellation at one half makes a
state-independent rate possible and separates the two parts of the spectrum.  For
the long-action moment costs considered here, this sign change echoes---but does not
resolve---the finite-horizon
R\'enyi-control distinctions at unit efficiency of
Ref.~\cite{TeoCombesWiseman2014}.

For a qubit, the pivot is sharper.  Evaluating the half-moment generator in
closed form [SM~\cite{SM}] shows that, for $\ell>0$ and $0<S<\tfrac12$, QND is,
up to unitary symmetries that leave the measurement generator invariant, its unique
maximizing orientation at every $\eta<1$, while at $S=\tfrac12$ all orientations
are instantaneously equivalent.  At $\eta=1$ the orientation dependence vanishes
entirely, so for \emph{every} admissible predictable protocol
\begin{equation}
\E_\pi[\sqrt{\det\rho_A}]
=\sqrt{\det\rho_0}\,e^{-2\ell^2A}.
\label{eq:invariant}
\end{equation}
Equation~\eqref{eq:invariant} suggests a sharp null test for future
near-unit-efficiency diffusive experiments that vary feedback at fixed action;
existing superconducting experiments separately demonstrate the requisite trajectory
reconstruction and fast-feedback capabilities~\cite{Murch2013,Campagne2013}.

\emph{A full-rank spectral-variance obstruction.---}
Rank-two boundary faces attain the uniform constants.  The determinant-root law
gives a strict full-rank obstruction whenever Eq.~\eqref{eq:detceiling} is smaller
than the uniform ceiling; for generic spectra this occurs over an explicit
moment-order interval.  Let
$R=(\det\rho)^{1/d}$ and use the fixed-spectrum invariants $V_L$ from
Eq.~\eqref{eq:spectralvariance} and $\mu=\Tr L/d$.  For a minimal purification,
$R$ is the G-concurrence divided by $d$.  At unit efficiency, each observed update
acts as a local filter, so Gour's covariance applies~\cite{Gour2005}.  Matrix It\^o
calculus gives, for every full-rank trajectory,
\begin{align}
\frac{dR}{R}
&=\left[-2\eta V_L+\frac{1-\eta}{d}K(\rho,L)\right]dA\notag\\[-2pt]
&\quad+2\sqrt\eta(\mu-\Tr\rho L)dW_A,
\label{eq:detrootSDE}\\
K(\rho,L)
&:=\Tr(\rho^{-1}L\rho L)-\Tr L^2\geq0.
\label{eq:Kpositive}
\end{align}
In the eigenbasis of $\rho$, $K$ is a sum of nonnegative terms [SM~\cite{SM}].
Hamiltonian control does not enter because it preserves the determinant.  The
bounded-noise localization in the SM preserves full rank at every finite action and
makes $e^{2\eta V_LA}R_A$ a true submartingale.  Hence
\begin{equation}
\E_\pi[(\det\rho_A)^{1/d}]
\geq(\det\rho_0)^{1/d}e^{-2\eta V_LA}.
\label{eq:detrootbound}
\end{equation}
At unit efficiency, this is the corresponding
protocol-independent determinant law.  With detection inefficiency, a fixed
commuting QND protocol attains the lower bound, whereas noncommuting dynamics
contributes through the nonnegative correction in Eq.~\eqref{eq:detrootSDE}.

A pointwise spectral comparison, $S^{\alpha_d}\geq[2(d-1)/d]^{\alpha_d}R$
with $\alpha_d=(d-1)/d$, followed by Jensen's inequality for $\theta>\alpha_d$
[SM~\cite{SM}], converts Eq.~\eqref{eq:detrootbound} into
Eq.~\eqref{eq:detceiling} at every efficiency.  For the canonical qutrit spectrum
$(-\ell,0,+\ell)$, $V_L=\tfrac23\ell^2$, so
$\Lambda_\pi\leq\tfrac43\eta\ell^2\max(1,\tfrac{3\theta}{2})$.  For $0<\eta<1$,
every full-rank qutrit initial state therefore remains strictly below the rank-two
ceiling throughout $(1-1/\sqrt3)/2<\theta<1$, irrespective of adaptive
noncommuting control.  More generally, $V_L=2\ell^2/d$ for
$L_d=\operatorname{diag}(-\ell,0^{d-2},+\ell)$.  The resulting strict interval
$[1-\sqrt{1-2/d}]/2<\theta<(d-1)/2$ is shown in Fig.~\ref{fig:freezing}(c).  Its
collapse at $d=2$ reflects the qubit identity $V_L=\ell^2$.  The
determinant bound is vacuous on a rank-two face because $R_0=0$.

\emph{The frozen branch: QND attainment and an all-dimensional ceiling.---}
The frozen branch follows from two complementary results.  The QND protocol attains
the asymptotic ceiling, and no admissible predictable protocol in any finite
dimension exceeds it for $0<\eta<1$.  In the aligned basis, the qubit remains
diagonal.  With the signed
population coordinate $y=\operatorname{arctanh}z$, where
$S=\tfrac12\operatorname{sech}^2y$, its diffusion is
\begin{equation}
dy=4\eta\ell^2\tanh y\,dA+2\ell\sqrt\eta\,dW_A.
\label{eq:y}
\end{equation}
The fixed-strength QND transition law is explicit~\cite{Li2013,Jiang2020}; the
measurement-action time change gives the propagator used here.  A ground-state
transformation reduces the evolution to free heat flow with an additive spectral
shift $2\eta\ell^2$ [SM~\cite{SM}].
For a maximally mixed initial state at unit efficiency, this propagator agrees,
after time normalization, with the Onsager--Machlup path-integral solution of
Ref.~\cite{Poltronieri2026}.  For
$S^\theta=2^{-\theta}\operatorname{sech}^{2\theta}y$, the Gaussian convolution of
variance $v=4\eta\ell^2A$ carries the weight $\cosh^{1-2\theta}y$ [SM~\cite{SM}].
Below the half moment, this weight grows at large $|y|$, and a moving Gaussian saddle
sets the exponent.  Above it, the weight is integrable, and the transformed spectral
edge determines the decay.  At the half moment, the weight is constant.  This change
of saddle gives the long-action spectrum
\begin{equation}
\Lambda_{\rm QND}(\theta,\eta)=
\begin{cases}
8\theta(1-\theta)\eta\ell^2,&0<\theta\leq\tfrac12,\\
2\eta\ell^2,&\theta\geq\tfrac12.
\end{cases}.
\label{eq:qndspectrum}
\end{equation}
For $\theta<\tfrac12$, the first line arises from a moving Gaussian saddle in the
measurement record.  For $\theta>\tfrac12$, the spectral edge produces an
$A^{-1/2}e^{-2\eta\ell^2A}$ tail.  At $\theta=\tfrac12$, this algebraic factor is
absent, and Eq.~\eqref{eq:rootbound} is saturated by a pure exponential.  Comparing
Eqs.~\eqref{eq:envelope} and \eqref{eq:qndspectrum} proves the first branch of
Eq.~\eqref{eq:phase}.  In large-deviation language, the moving QND saddle reaches
the physical boundary $-A^{-1}\ln S_A=0$ at $\theta=\tfrac12$ and remains there,
forming a terminated, moment-independent
branch~\cite{Touchette2009,Fyodorov2009,DemboZeitouni1998}.

The SM gives the near-pure rigidity estimate
$T_2\leq12\ell^2S^2+8S\Delta$, with $\Delta=T_1-T_2\geq0$, once the largest
eigenvalue is isolated.  Hence the drift coefficient
$B=2\eta T_2-(1-\eta)\Delta$ in $dS=-2B\,dA-4\sqrt\eta\,T_3dW_A$ obeys
$B\leq24\eta\ell^2S^2$ on the corridor
$S\leq s_\eta=\min\{\tfrac14,(1-\eta)/(16\eta)\}$, which is nonempty precisely for
$0<\eta<1$.  Along a near-pure direction with insufficient innovation noise, $S$
can then decrease only algebraically, rather than at a positive exponential
rate.  A change of
measure for $X=\tfrac12\ln S$ on the corridor $\{S\leq s\}$ cancels the tilted drift
at a reverse-likelihood cost of at most $2\eta\ell^2A$~\cite{KaratzasShreve1991}.
Together with a last-excursion estimate and the continuous-martingale maximal
inequality, this gives
\begin{equation}
\limsup_{A\to\infty}-A^{-1}\ln\E_\pi[S_A^\theta]\leq2\eta\ell^2,
\label{eq:alldlimsup}
\end{equation}
for every $0<\eta<1$, every fixed $\theta>0$, every mixed initial state, and
every admissible predictable protocol held fixed as $A\to\infty$, in every finite
dimension [SM~\cite{SM}].  Below
$\theta=\tfrac12$, the finite-action bound \eqref{eq:envelope} is stronger; together
they give Eq.~\eqref{eq:alldceiling}.

The half-moment kink and the singular efficiency endpoint have distinct
finite-action resolutions, derived in the End Matter.

\emph{The freezing mechanism and the singular endpoint.---}
A qubit admits a stronger event-level explanation than the all-dimensional
ceiling.  Fix $0<\eta<1$ and $\omega\in(0,\sqrt{1-\eta}]$ small enough that
$s_Y<S_0$, where $\psi_\omega$ is the localized comparison function and $s_Y$
the impurity threshold constructed in the SM.  Stopping the submartingale
$e^{\lambda_\omega A}\psi_\omega(S_A)$, with
$\lambda_\omega=2\eta\ell^2(1+\omega^2)$, when $S$ first reaches $s_Y$ gives,
for every $\theta>0$ and every admissible predictable qubit protocol,
\begin{equation}
\E_\pi[S_A^\theta]\geq s_Y^\theta\psi_\omega(S_0)
e^{-2\eta\ell^2(1+\omega^2)A}.
\label{eq:survival}
\end{equation}
The positive prefactor is independent of $A$.  Taking first $A\to\infty$ at fixed
$\omega$ and only then $\omega\downarrow0$ recovers the frozen ceiling.  The
underlying survival certificate shows that a nonvanishing exponentially weighted
probability remains above a fixed impurity threshold.  Thus persistently mixed
records, rather than typical trajectories, determine the frozen branch.  The
certificate holds for every admissible predictable control law; no stationarity or
Markov assumption is imposed.

For unbiased feedback the impurity is instead deterministic,
\begin{equation}
S_A=\frac{1-\eta}{2}
+\left(S_0-\frac{1-\eta}{2}\right)e^{-4\ell^2A}.
\label{eq:unbiased}
\end{equation}
At unit efficiency, this solution attains the upper branch
$4\theta\ell^2$ of the envelope~\eqref{eq:envelope}.  For every $\eta<1$, it
instead stalls at $(1-\eta)/2$, and the admissible window for $\omega$ collapses
precisely at $\eta=1$.  Although a controller may switch bases before the stall,
the protocol-uniform bound~\eqref{eq:survival} excludes an improvement of the
asymptotic exponent.  Finite-horizon policies may nevertheless switch between QND
and unbiased orientations, consistent with the finite-time landscape under detection
inefficiency in
Ref.~\cite{Jiang2020}.  Hence, for every $\theta>\tfrac12$,
\begin{equation}
\lim_{\eta\to1^-}\Lambda_\star(\theta,\eta)=2\ell^2
\neq4\theta\ell^2=\Lambda_\star(\theta,1).
\label{eq:singular}
\end{equation}
The limit is nonuniform.  The two fixed-policy finite-action laws remain continuous
in $\eta$, and the singularity appears only after taking the long-action limit.
The late QND--unbiased crossing is delayed logarithmically as $\eta\to1^-$, as
quantified in the End Matter and SM~\cite{SM}; within this fixed-endpoint
diagnostic, finite-action observations can therefore appear unit-efficient long
before the frozen tail emerges.

\emph{Discussion.---}
Detection inefficiency does more than rescale an ideal purification rate.  Unbiased
feedback suppresses trajectory fluctuations but stalls at a nonzero impurity floor,
whereas QND purification continues toward purity while retaining exponentially rare,
persistently mixed records.  The protocol-uniform survival bound shows that no
admissible predictable protocol can eliminate the high-order contribution of such
records; QND makes that contribution rate determining and attains the resulting
bound.  Because every strictly mixed qubit state has rank two, this yields the exact
frozen optimum for every $0<\eta<1$.  In higher dimensions, the full-rank optimum
remains open; within the proven gap, only strict separation from the rank-two
constant is known.  Balanced dichotomic spectra and extensions to jump records remain
important cases.  The
boundary--interior separation can be tested at fixed action by comparing full-rank
qutrit states with embedded two-level mixtures, while added record noise controls
the effective detection efficiency $\eta$.

\emph{Data availability.---} No external data were used.  All analytic results and
numerical evaluations are contained in the Letter and Supplemental Material.

\clearpage
\onecolumngrid

\setcounter{equation}{0}
\renewcommand{\theequation}{A\arabic{equation}}
\renewcommand{\theHequation}{A\arabic{equation}}
\section*{End Matter}
\begin{figure}[!htb]
\centering
\includegraphics[width=0.985\textwidth]{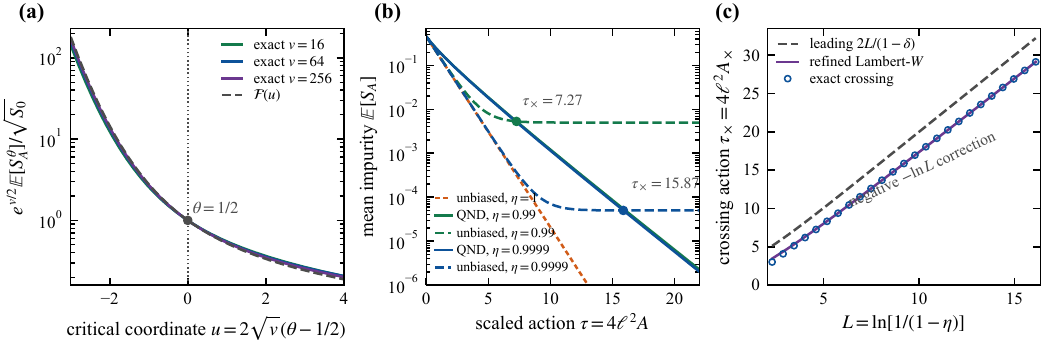}
\caption{Finite-action resolution of the two asymptotic singularities at
$S_0=0.455$.  (a) Fixed-QND moments collapse onto the scaling function
$\mathcal F(u)$ in Eq.~\eqref{eq:critical-layer}.
(b) Exact QND mean impurity (solid) and always-unbiased flows (dashed) near unit
efficiency; circles mark the logarithmically delayed crossings.
(c) Exact crossing actions follow the refined Lambert-$W$ law in
Eq.~\eqref{eq:endmatter-crossing}, including the negative $-\ln L$ correction;
the leading $2L/(1-\delta)$ estimate converges slowly.  Curves are computed by
converged numerical quadrature of the exact QND kernel; no trajectory time
discretization is used.}
\label{fig:endmatter}
\end{figure}

\noindent\begin{minipage}[t]{0.485\textwidth}
\vspace{0pt}
\emph{Finite-action resolution of the two singular limits.---}%
The asymptotic laws contain two boundary layers: one in moment order and one in
efficiency--action space.  Both are already visible in the exact fixed-endpoint
protocols, without fitting a trajectory-discretized model.

\emph{Moment-order layer.---}
Fix $0<\eta\leq1$, $\ell>0$, and a mixed aligned initial state
$S_0=\tfrac12\operatorname{sech}^2y_0>0$.  With
$v=4\eta\ell^2A\to\infty$ and
$\theta=\tfrac12+u/(2\sqrt v)$, the exact heat kernel [SM~\cite{SM}] gives,
uniformly for $u$ in compact sets,
\begin{align}
\frac{e^{v/2}\E_{\rm QND}[S_A^\theta]}{\sqrt{S_0}}
&=2^{-u/(2\sqrt v)}
\E\!\left[\cosh(y_0+\sqrt v Z)^{-u/\sqrt v}\right]\notag\\
&\longrightarrow
\mathcal F(u):=\E[e^{-u|Z|}]\notag\\
&=e^{u^2/2}\operatorname{erfc}\!\left(\frac{u}{\sqrt2}\right),
\label{eq:critical-layer}
\end{align}
where $Z\sim N(0,1)$; the convergence follows from
$v^{-1/2}\ln\cosh(y_0+\sqrt vZ)\to|Z|$ and a Gaussian-integrable bound on compact
$u$-sets.  The leading profile is parameter free: $y_0$ enters only preasymptotic
corrections, while $\eta$ and $\ell$ only set $v$.  The limits $\mathcal F(0)=1$,
$\mathcal F(u)\sim\sqrt{2/\pi}/u$ as $u\to+\infty$, and
$\mathcal F(u)\sim2e^{u^2/2}$ as $u\to-\infty$ match the exact half moment, the
frozen spectral edge, and the two moving saddles, respectively.  Thus
$\Delta\theta=O(v^{-1/2})=O[(\eta\ell^2A)^{-1/2}]$.  At finite action,
$|u|\lesssim1$ identifies the rounded layer.  Moments at several actions can be
tested for the normalized collapse in Fig.~\ref{fig:endmatter}(a), rather than
fitted to a discontinuous exponent at finite $A$.  At fixed $\theta\ne\tfrac12$,
$u\to\pm\infty$ recovers the two outer branches.
\end{minipage}\hfill
\begin{minipage}[t]{0.485\textwidth}
\vspace{0pt}

\emph{Efficiency--action layer.---}
At $\theta=1$ put $\delta=1-\eta$.  The same kernel and the always-unbiased
solution give, with $B=\tfrac12(y_0^2+\pi^2/4)$,
\begin{align}
\E_{\rm QND}[S_A]
&=\frac{\sqrt{\pi S_0}}{2\sqrt v}e^{-v/2}
  [1-B/v+O(v^{-2})],\notag\\
S_A^{\rm ub}
&=\frac{\delta}{2}+\left(S_0-\frac{\delta}{2}\right)e^{-v/\eta}.
\label{eq:endmatter-laws}
\end{align}
At the nonzero late crossing, with $S_0$ fixed and $\delta\downarrow0$, the
unbiased transient is $o(\delta)$.  With $L=\ln(1/\delta)$ and
$W_\delta=W_0(\pi S_0/\delta^2)$, where $W_0$ denotes the principal branch of
the Lambert $W$ function, its action obeys
\begin{align}
v_\times:=4\eta\ell^2A_\times
&=W_\delta-\frac{2B}{1+W_\delta}+O(W_\delta^{-2})\notag\\
&=2L-\ln(2L)+\ln(\pi S_0)
  +O\!\left(\frac{\ln L}{L}\right).
\label{eq:endmatter-crossing}
\end{align}
With the inner coordinate $w=v-2L+\ln(2L)-\ln(\pi S_0)$ held fixed as
$\delta\downarrow0$,
\begin{equation}
\left(\frac{\E_{\rm QND}[S_A]}{\delta/2},
      \frac{S_A^{\rm ub}}{\delta/2}\right)
\longrightarrow (e^{-w/2},1).
\label{eq:endmatter-efficiency-layer}
\end{equation}
The late fixed-endpoint crossing lies within a window $v\leq v_{\max}$ only if
$v_{\max}\gtrsim W_0(\pi S_0/\delta^2)$, or, equivalently at leading order,
$\delta\gtrsim[\pi S_0/(v_{\max}e^{v_{\max}})]^{1/2}$.
Below that resolvable loss, the crossover remains unresolved within the observation
window.  These fixed-endpoint diagnostics do not establish finite-horizon optimality
for switching policies.
\end{minipage}

\clearpage
\onecolumngrid
\setcounter{secnumdepth}{3}
\setcounter{section}{0}
\setcounter{subsection}{0}
\setcounter{equation}{0}
\setcounter{figure}{0}
\setcounter{table}{0}
\setcounter{footnote}{0}
\renewcommand{\theequation}{S\arabic{equation}}
\renewcommand{\theHequation}{SM.S\arabic{equation}}
\renewcommand{\thesection}{\Roman{section}}
\renewcommand{\theHsection}{SM.\arabic{section}}
\renewcommand{\thesubsection}{\thesection.\Alph{subsection}}
\renewcommand{\theHsubsection}{SM.\arabic{section}.\arabic{subsection}}
\renewcommand{\thetable}{S\Roman{table}}
\renewcommand{\theHtable}{SM.S\arabic{table}}
\renewcommand{\thefigure}{S\arabic{figure}}
\renewcommand{\theHfigure}{SM.S\arabic{figure}}

\begin{center}
{\large\bfseries Supplemental Material for\\[3pt]
Efficiency-Induced Freezing in Quantum-State Purification}

\end{center}
\medskip

This Supplemental Material gives the qubit spectral parametrization and endpoint
linearity (Sec.~\ref{sm:param}); the general-dimensional trace frame, variance and
quartic inequalities, quartic equality classification, and master inequality
(Sec.~\ref{sm:ddim}); the full-rank determinant law and its spectral-variance
obstruction (Sec.~\ref{sm:detlaw}); the qubit
half-moment pivot law (Sec.~\ref{sm:sqrt}); the finite-action moment envelope
(Secs.~\ref{sm:ddimmoment} and \ref{sm:moments}); the dimension-independent
inefficient moment ceiling (Sec.~\ref{sm:alldfreeze}); the exact QND heat kernel,
large-deviation form, critical moment-order layer, and late protocol crossover
(Sec.~\ref{sm:exact}); the first-passage corollary and exact finite-threshold QND
formula (Sec.~\ref{sm:fp}); the qubit fixed-threshold certificates and
moment-ceiling attainment proof (Sec.~\ref{sm:frozen}); and reproducibility details for the figure
and numerical stress tests (Sec.~\ref{sm:numerics}).

\section{Spectral parametrization and linearity in the measurement basis}
\label{sm:param}

\subsection{Invariants}

Write the qubit state in its instantaneous eigenbasis, $\rho=\mathrm{diag}(\lambda,
1-\lambda)$ with $\lambda\ge\tfrac12$, impurity $S=1-\Tr\rho^{2}=2\lambda(1-\lambda)
=2\det\rho\in[0,\tfrac12]$, and gap parameter $\beta=(2\lambda-1)^{2}=1-2S$. The
measured observable $L$ is Hermitian with eigenvalues $\pm\ell$; in the state
eigenbasis its matrix elements satisfy the basis-independent identity
\begin{equation}
|L_{12}|^{2}+\Delta^{2}/4=\ell^{2},
\qquad \Delta \equiv L_{11}-L_{22},
\label{eq:sm-identity}
\end{equation}
because $L^{2}=\ell^{2}\openone$ for a traceless Hermitian qubit observable (a trace
part shifts $L$ by a multiple of the identity and drops out of every expression below
through the centering $\tilde L=L-\Tr(L\rho)$). The relative orientation of $L$ and
$\rho$ therefore enters through the single variable
\begin{equation}
x \equiv \Delta^{2}/4 \in [0,\ell^{2}],
\end{equation}
with $x=\ell^{2}$ the aligned (QND) basis and $x=0$ the unbiased basis. Direct
evaluation in the eigenbasis gives the three invariants used below,
\begin{align}
\Tr(\tilde L\rho\tilde L\rho) &= S\,(\ell^{2}-x\beta),
\label{eq:sm-inv1}\\
\Tr(\tilde L^{2}\rho^{2}) &= (\ell^{2}-x)(1-S)+2xS^{2},
\label{eq:sm-inv2}\\
\big[\Tr(\tilde L\rho^{2})\big]^{2} &= x\,S^{2}\beta .
\label{eq:sm-inv3}
\end{align}
For Eq.~(\ref{eq:sm-inv1}), $\tilde L_{11}=(1-\lambda)\Delta$,
$\tilde L_{22}=-\lambda\Delta$, so the diagonal part contributes
$2\lambda^{2}(1-\lambda)^{2}\Delta^{2}=2xS^{2}$ and the off-diagonal part
$2\lambda(1-\lambda)|L_{12}|^{2}=S(\ell^{2}-x)$; combining with
Eq.~(\ref{eq:sm-identity}) yields $S[\ell^{2}-x(1-2S)]=S(\ell^{2}-x\beta)$.
Equations~(\ref{eq:sm-inv2}) and (\ref{eq:sm-inv3}) follow the same way.

\subsection{Impurity dynamics}

It\^o's formula for $Q=\Tr\rho^{2}$ under Eq.~(1) of the Letter gives
$dQ=2\Tr(\rho\,d\rho)+\Tr[(d\rho)^{2}]$. The Hamiltonian term contributes
$2\Tr(\rho[H,\rho])=0$; more generally, any control acting by unitary conjugation
preserves the spectrum of $\rho$ pathwise and cannot enter the dynamics of $S$. The
measurement terms give, using the cyclic identities
$2\Tr(\rho\,\mathcal{D}[L]\rho)=2[\Tr(\tilde L\rho\tilde L\rho)-\Tr(\tilde
L^{2}\rho^{2})]$ and $\Tr\{(\mathcal{H}[L]\rho)^{2}\}
=2\Tr(\tilde L\rho\tilde L\rho)+2\Tr(\tilde L^{2}\rho^{2})$
(both invariant under the centering, which cancels between the two expansions),
\begin{equation}
\begin{aligned}
dS &= -2M\,g(x,S)\,dt - 4\sqrt{\eta M\,x\beta}\;S\,dW ,\\
g(x,S) &= (\eta-1)\big[(\ell^{2}-x)(1-S)+2xS^{2}\big]+(1+\eta)\,S\,(\ell^{2}-x\beta),
\end{aligned}
\label{eq:sm-sde}
\end{equation}
where the diffusion coefficient is $2\sqrt{\eta M}\cdot 2|\Tr(\tilde L\rho^{2})|$ by
Eq.~(\ref{eq:sm-inv3}). Both the drift and the squared diffusion of $S$ (and hence
every generator appearing in this problem) are \emph{affine in $x$}. Consequently, for
any functional $F(S)$ the supremum or infimum of $\mathcal{A}F$ over the measurement
basis is attained at an endpoint $x\in\{0,\ell^{2}\}$; pointwise generator
optimization therefore reduces to the unbiased and the aligned bases, and every
global statement below is proved through such endpoint generator inequalities. With adaptive basis control,
$x_t$ is a predictable (nonanticipating) process with values in $[0,\ell^{2}]$;
every bound below is a pointwise bound on the generator, hence holds for arbitrary
such $x_t$, finite-variation unitary control $H_t$, and scheduled strength $M(t)$.
Measurement-current-proportional stochastic Hamiltonian feedback is not part of this
resource model.
The same reduction to the aligned and unbiased endpoints appears in the finite-horizon
mean-Bloch-length problem of Jiang \emph{et al.}~\cite{Jiang2020}; the cost and the
asymptotic theorem studied here are different.

At the unbiased endpoint the diffusion vanishes, $g(0,S)=\ell^{2}[(1+\eta)S-(1-\eta)(1-S)]$,
and the impurity flows deterministically,
\begin{equation}
\dot S=-2M\ell^{2}\big[(1+\eta)S-(1-\eta)(1-S)\big],
\label{eq:sm-ubbflow}
\end{equation}
which vanishes at $S_{\rm stall}=(1-\eta)/2$: for $\eta<1$ the unbiased basis cannot
purify below this floor, as quoted in the Letter.

\section{General-dimension framework}
\label{sm:ddim}

The Letter states its laws for a state $\rho$ of arbitrary finite dimension
$d$ and an observable of fixed spectral half-spread $\ell$; the qubit of
Sec.~\ref{sm:param} is the exactly-solvable face. This section builds the
general-$d$ apparatus: a three-invariant trace frame, the master inequality
$C_{\sqrt S}$ that carries the $\sqrt S$ speed limit, and its proof through a variance
bound (Lemma~A), a freshly established quartic inequality (Sec.~\ref{sm:ct2}), and
a one-line efficiency reduction (Lemma~R).

\subsection{The trace frame and the impurity equation}

Let $L$ be Hermitian with spectrum contained in an interval of width $2\ell$
(half-spread $\ell$), $\tilde L=L-\Tr(\rho L)$ the centered observable, and
$S=1-\Tr\rho^{2}$ the impurity. The relative orientation of $\rho$ and $\tilde L$
enters every generator below only through the three real invariants
\begin{equation}
T_{1}=\Tr(\tilde L^{2}\rho^{2}),\qquad
T_{2}=\Tr(\tilde L\rho\tilde L\rho),\qquad
T_{3}=\Tr(\rho^{2}\tilde L),
\label{sm:ddim-frame}
\end{equation}
whose qubit values are Eqs.~(\ref{eq:sm-inv2}), (\ref{eq:sm-inv1}) and
(\ref{eq:sm-inv3}): $T_{1}=(\ell^{2}-x)(1-S)+2xS^{2}$, $T_{2}=S(\ell^{2}-x\beta)$,
$T_{3}^{2}=xS^{2}\beta$. It\^o's formula for $Q=\Tr\rho^{2}$ under Eq.~(1),
together with the commutator identity
$\|[\tilde L,\rho]\|_{F}^{2}=2(T_{1}-T_{2})$, gives the dimension-free impurity
equation
\begin{equation}
dS=-2M\big[(1+\eta)T_{2}-(1-\eta)T_{1}\big]\,dt
\;-\;4\sqrt{\eta M}\,T_{3}\,dW,
\qquad (dS)^{2}=16\,\eta M\,T_{3}^{2}\,dt.
\label{sm:ddim-sde}
\end{equation}
Unitary control preserves the spectrum of $\rho$ and cannot enter; only
$(T_{1},T_{2},T_{3})$ do. Every bound below is pointwise in the state, hence holds
for every predictable measurement strength $M(t)$ and basis, and every
finite-variation Hamiltonian control $H(t)$.
Applying It\^o's formula to $\sqrt S$ with Eq.~(\ref{sm:ddim-sde}) gives
\begin{equation}
-\frac{\mathcal{A}\sqrt S}{\sqrt S}
=M\Big[\,\frac{(1+\eta)T_{2}-(1-\eta)T_{1}}{S}
+2\eta\,\frac{T_{3}^{2}}{S^{2}}\,\Big],
\label{sm:ddim-sqrtgen}
\end{equation}
the $d$-dimensional counterpart of Eq.~(\ref{eq:sm-sqrtgen}). The
pointwise $\sqrt S$ speed limit $-\mathcal{A}\sqrt S\le2\eta\ell^{2}M\sqrt S$ is
therefore, for every state with $S>0$, equivalent to the \emph{master inequality}
\begin{equation*}
(1+\eta)T_{2}-(1-\eta)T_{1}+2\eta\,\frac{T_{3}^{2}}{S}
\;\le\;2\eta\ell^{2}S,
\tag{$C_{\sqrt S}$}
\end{equation*}
which we now establish for all $d$, every state with $S>0$, all centered Hermitian
$\tilde L$ of half-spread $\ell$, and all $\eta\in[0,1]$
[labeled Eq.~(\ref{sm:csqrt}) at its point of proof].
At the pure boundary the multiplied form
$(1+\eta)T_{2}S-(1-\eta)T_{1}S+2\eta T_{3}^{2}\le2\eta\ell^{2}S^{2}$ holds
trivially (both sides vanish at $S=0$, where $T_{3}=\Tr(\rho\widetilde L)=0$ for a
pure state), and the half-moment bound extends by stopping at $\{S=\epsilon\}$ and
continuity. At $d=2$ the quartic
ingredient is the identity $T_{2}S+T_{3}^{2}=\ell^{2}S^{2}$ for every basis.
For $\eta<1$, equality in the resulting $C_{\sqrt S}$ bound additionally
requires the aligned basis (apart from $S=\tfrac12$); for $\eta=1$ every qubit
basis is an equality case, as shown directly in Sec.~\ref{sm:sqrt}.

\subsection{Lemma~A: a variance bound on \texorpdfstring{$T_{3}$}{T3}}

\textbf{Lemma~A.}
\label{sm:lemA}
For every state and every centered Hermitian $\tilde L$ of half-spread
$\ell$,
\begin{equation}
|T_{3}|=\big|\Tr(\rho^{2}\tilde L)\big|\;\le\;\ell\,(1-\Tr\rho^{2})=\ell\,S.
\end{equation}
\emph{Proof} (four lines). Work in the eigenbasis $\rho=\mathrm{diag}(\lambda_{i})$
and let $m_{i}=\tilde L_{ii}$ be the diagonal of $\tilde L$; centering reads
$\sum_{i}\lambda_{i}m_{i}=0$ and the spectral window is $m_{i}\in[u,u+2\ell]$ for
some $u$. Put $\tilde m_{i}=m_{i}-(u+\ell)\in[-\ell,\ell]$ and
$s=\sum_{i}\lambda_{i}\tilde m_{i}=-(u+\ell)$; with $r_{2}=\Tr\rho^{2}=\sum_{i}\lambda_{i}^{2}$,
since $\rho^{2}$ is diagonal only $m_{i}$ enters and
\begin{equation}
T_{3}=\sum_{i}\lambda_{i}^{2}m_{i}
=\sum_{i}\lambda_{i}^{2}\tilde m_{i}-s\,r_{2}
=\sum_{i}\lambda_{i}(\lambda_{i}-r_{2})\,\tilde m_{i},
\end{equation}
the $s$-term cancelling identically. Pointwise $|\lambda_{i}-r_{2}|\le
1-\lambda_{i}$ in both signs: $r_{2}-\lambda_{i}\le1-\lambda_{i}$ because
$r_{2}\le1$, and $\lambda_{i}-r_{2}\le1-\lambda_{i}$ because
$\lambda_{i}^{2}\le r_{2}$ gives $2\lambda_{i}\le1+\lambda_{i}^{2}\le1+r_{2}$.
Hence $|T_{3}|\le\ell\sum_{i}\lambda_{i}(1-\lambda_{i})
=\ell(1-\sum_{i}\lambda_{i}^{2})=\ell S$. $\blacksquare$
The qubit case is the identity $T_{3}^{2}=xS^{2}\beta\le\ell^{2}S^{2}$.

\subsection{The quartic inequality \texorpdfstring{$C_{T2}'$}{C-T2 prime}
and its equality cases}

\textbf{Theorem ($C_{T2}'$).}
\label{sm:ct2}
Let $\rho$ be a finite-dimensional state, let $L$ be Hermitian with spectral
half-spread $\ell$, and let $P_\rho$ denote the support projector of $\rho$.
The width of $P_\rho L P_\rho$ below is understood as the width of its
compression to $\operatorname{supp}\rho$.
Then
\begin{equation}
T_{2}S+T_{3}^{2}\leq \ell^{2}S^{2}.
\label{eq:sm-ct2}
\end{equation}
If $S=0$, equality holds trivially; if $\ell=0$, equality holds for every
state.  If $S>0$ and $\ell>0$, equality holds if and only if
\begin{equation}
\operatorname{rank}\rho=2,
\qquad
\operatorname{width}(P_\rho L P_\rho)=2\ell .
\label{eq:sm-ct2-equality}
\end{equation}
Equivalently, the support is spanned by one vector from each of the minimum-
and maximum-eigenvalue eigenspaces of $L$.

\emph{Proof.}
Work in an eigenbasis
$\rho=\operatorname{diag}(\lambda_1,\ldots,\lambda_d)$ and, for
$t\in\mathbb R$, put $N_t=\widetilde L+t\openone$.  Since
$\Tr(\rho\widetilde L)=0$,
\begin{align}
&T_2+2tT_3-t^2S \nonumber\\
&\quad =
\Tr(\rho N_t\rho N_t)-[\Tr(\rho N_t)]^2 \nonumber\\
&\quad =
2\sum_{i<j}\lambda_i\lambda_j
\left(|(N_t)_{ij}|^2-(N_t)_{ii}(N_t)_{jj}\right) \nonumber\\
&\quad =
2\sum_{i<j}\lambda_i\lambda_j
\big[-\det B_{ij}(t)\big],
\label{eq:sm-pairdet}
\end{align}
where $B_{ij}(t)$ is the $2\times2$ principal compression of $N_t$ to the
span of the $i$th and $j$th eigenvectors of $\rho$.

Write the eigenvalues of $B_{ij}(t)$ as $c_{ij}\pm d_{ij}$.  Compression
cannot increase spectral width, while scalar shifts leave it unchanged, so
$0\leq d_{ij}\leq\ell$.  Consequently,
\begin{equation}
-\det B_{ij}(t)=d_{ij}^{2}-c_{ij}^{2}\leq\ell^2 .
\end{equation}
Using $S=1-\sum_i\lambda_i^2=2\sum_{i<j}\lambda_i\lambda_j$ in
Eq.~\eqref{eq:sm-pairdet} gives, for every $t$,
\begin{equation}
T_2+2tT_3-t^2S\leq\ell^2S .
\end{equation}
For $S>0$, maximizing the left-hand side at $t_\star=T_3/S$ yields
$T_2+T_3^2/S\leq\ell^2S$, which is Eq.~\eqref{eq:sm-ct2}.

It remains to classify equality.  Assume $S>0$ and $\ell>0$.  Equality in
Eq.~\eqref{eq:sm-ct2} requires equality in every positive-weight summand of
Eq.~\eqref{eq:sm-pairdet} at $t=t_\star$.  Hence, for every $i<j$ with
$\lambda_i\lambda_j>0$, the two eigenvalues of $B_{ij}(t_\star)$ must be
exactly $\pm\ell$.

Suppose that $\operatorname{rank}\rho\geq3$ and choose three indices $i,j,k$
in its support.  The trace-zero condition for the three pair compressions gives
\begin{equation*}
(N_{t_\star})_{ii}=(N_{t_\star})_{jj}=(N_{t_\star})_{kk}=0,
\end{equation*}
and their determinant conditions give
\begin{equation*}
|(N_{t_\star})_{ij}|=|(N_{t_\star})_{ik}|
=|(N_{t_\star})_{jk}|=\ell .
\end{equation*}
The corresponding $3\times3$ compression $C$ therefore satisfies
$\Tr C=0$ and $\Tr C^2=6\ell^2$.  On the other hand,
$\operatorname{width}(C)\leq2\ell$, so the elementary variance bound for its
three real eigenvalues gives
\begin{equation}
\frac{1}{3}\Tr C^2
\leq\frac{\operatorname{width}(C)^2}{4}
\leq\ell^2,
\end{equation}
contradicting $\Tr C^2/3=2\ell^2$.  Thus equality requires
$\operatorname{rank}\rho=2$.

Let the two nonzero eigenvalues of $\rho$ be $p,q>0$, $p+q=1$, and let $a,b$
be the corresponding diagonal entries of $\widetilde L$.  Centering gives
$pa+qb=0$, while $S=2pq$ and $T_3=p^2a+q^2b$.  Therefore
\begin{equation}
t_\star=\frac{T_3}{S}=-\frac{a+b}{2}.
\end{equation}
Thus the support compression of $N_{t_\star}$ is traceless.  It has eigenvalues
$\pm\ell$ precisely when
$\operatorname{width}(P_\rho L P_\rho)=2\ell$, proving both necessity and
sufficiency in Eq.~\eqref{eq:sm-ct2-equality}.  Since the compression then reaches
both global spectral endpoints, equality in the Rayleigh principle places its two
extremal vectors in the corresponding extremal eigenspaces of $L$.  The cases
$S=0$ and $\ell=0$ follow directly from $T_2=T_3=0$. $\blacksquare$

At $d=2$ and $S>0$, $P_\rho=\openone$ and
$\operatorname{width}(L)=2\ell$, so Eq.~\eqref{eq:sm-ct2} is an identity for
every relative orientation of $\rho$ and $L$.

\textit{Perfect-efficiency corollary.}
At $\eta=1$, Eq.~\eqref{sm:ddim-sqrtgen} becomes
\begin{equation}
-\frac{\mathcal A\sqrt S}{\sqrt S}
=2M\,\frac{T_2S+T_3^2}{S^2}
\leq2\ell^2M .
\label{eq:sm-perfect-dichotomy}
\end{equation}
For $S>0$ and $\ell>0$, equality holds if and only if
$\operatorname{rank}\rho=2$ and
$\operatorname{width}(P_\rho L P_\rho)=2\ell$.  Thus every mixed qubit
saturates the pointwise ceiling for every measurement orientation, whereas
every state of support rank at least three is pointwise strict.  This is a
pointwise dichotomy: because full-rank states may approach an equality face
arbitrarily closely, it does not furnish a uniform higher-dimensional
long-action exponent gap.

\begin{figure}[t]
\centering
\includegraphics[width=0.92\textwidth]{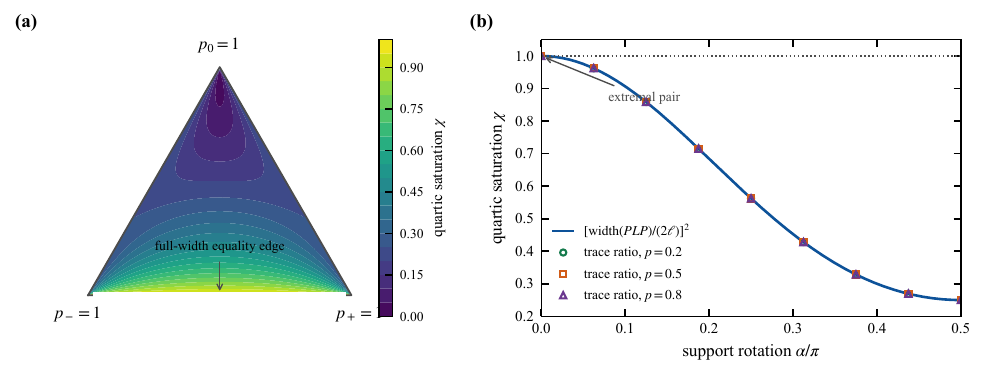}
\caption{Exact geometry of the quartic equality set for
$L=\ell\,\mathrm{diag}(-1,0,1)$.
(a) On the commuting diagonal slice
$\rho=\mathrm{diag}(p_-,p_0,p_+)$, the color is the saturation fraction
$\chi=(T_2S+T_3^2)/(\ell^2S^2)$.  The three pure vertices, where this normalized
ratio is $0/0$, are omitted.  Among mixed states only the extremal edge $p_0=0$
saturates; the two other rank-two edges have $\chi=1/4$, and the maximally mixed
state has $\chi=1/3$.
(b) For the noncommuting rank-two support spanned by
$|{-}\rangle$ and $\cos\alpha|{+}\rangle+\sin\alpha|0\rangle$,
$\chi=[(1+\cos^2\alpha)/2]^2$, independently of its two nonzero eigenvalues
$p$ and $1-p$.  Symbols are direct trace evaluations and the curve is the support
compression width.}
\label{fig:sm-equality}
\end{figure}

Figure~\ref{fig:sm-equality} makes the equality classification explicit for the
canonical qutrit.  It separates the condition ``rank two'' from the stronger
condition that the support compression span the full spectral width.

\subsection{Lemma~R: reduction of \texorpdfstring{$C_{\sqrt S}$}{C-sqrt-S} to
\texorpdfstring{$C_{T2}'$}{C-T2 prime}}

\textbf{Lemma~R.}
\label{sm:lemR}
Define
\begin{equation}
P:=(T_{1}-T_{2})\,S,\qquad
Q:=(T_{1}+T_{2})\,S+2T_{3}^{2}-2\ell^{2}S^{2}.
\end{equation}
Then $P=\tfrac12\|[\tilde L,\rho]\|_{F}^{2}\,S\ge0$ (Cauchy--Schwarz), and
for every $\eta\in[0,1]$ the master inequality $C_{\sqrt S}$ is equivalent to
$\eta Q\le P$. Indeed, multiplying $C_{\sqrt S}$ by $S$ and collecting, the excess is
\begin{equation}
\begin{aligned}
&(1+\eta)T_{2}S-(1-\eta)T_{1}S+2\eta T_{3}^{2}-2\eta\ell^{2}S^{2}\\
&\quad=-(T_{1}-T_{2})S+\eta\big[(T_{1}+T_{2})S+2T_{3}^{2}-2\ell^{2}S^{2}\big]
=-P+\eta Q,
\end{aligned}
\end{equation}
so $C_{\sqrt S}\Leftrightarrow\eta Q\le P$. Now $Q\le P$ is \emph{precisely}
$C_{T2}'$, since $Q\le P\Leftrightarrow2T_{2}S+2T_{3}^{2}\le2\ell^{2}S^{2}$. Hence,
once $C_{T2}'$ holds, for every $\eta\in[0,1]$
\begin{equation}
\eta Q\le\eta P\le P\ \ (\text{if }Q\ge0),
\qquad \eta Q\le0\le P\ \ (\text{if }Q<0),
\end{equation}
so $\eta Q\le P$ at every efficiency. The single, $\eta$-independent
inequality $C_{T2}'$ therefore implies the master inequality
\begin{equation}
(1+\eta)T_{2}-(1-\eta)T_{1}+2\eta\,\frac{T_{3}^{2}}{S}
\;\le\;2\eta\ell^{2}S
\label{sm:csqrt}
\end{equation}
for all $\eta\in[0,1]$, and with it, via Eq.~(\ref{sm:ddim-sqrtgen}), the
pointwise $\sqrt S$ speed limit $-\mathcal{A}\sqrt S\le2\eta\ell^{2}M\sqrt S$ in
every dimension: the $d$-dimensional master law behind the $\sqrt S$ theorem of
the Letter, of which the qubit $\sqrt{\det\rho}$ law (Sec.~\ref{sm:sqrt}) is the
equality face. $\blacksquare$

\section{Full-rank determinant law and spectral-variance obstruction}
\label{sm:detlaw}

This section proves that the rank-two sharpness construction cannot generically be
continued into the full-rank interior.  The result uses the fixed spectral multiset
of the measured observable, not only its half-spread.  Put
\begin{equation}
\mu=\frac{\Tr L}{d},\qquad
V_L=\frac1d\Tr(L-\mu\openone)^2,
\qquad R=(\det\rho)^{1/d}.
\label{eq:sm-VL}
\end{equation}
Both $\mu$ and $V_L$ are unchanged by the predictable unitary orientations allowed
in the Letter.  A scalar shift of $L$ leaves the SME unchanged, so in the
calculation below we first take $\mu=0$; the invariant form is restored at the end.
For a minimal $d\times d$ purification, $dR$ is the G-concurrence associated with its
Schmidt spectrum.  Gour established its covariance under local filtering
\cite{Gour2005}; adjacent one-sided-channel factorization laws for pure inputs, with
mixed-input bounds, were proved in Refs.~\cite{Konrad2008,Tiersch2008,Gour2010}.  Related
state-factorized mean-concurrence laws hold in local, unit-efficiency monitoring of
Markovian two-qubit reservoir models
\cite{Viviescas2010,VogelsbergerSpehner2010,GuevaraViviescas2014}.  The lossless determinant factorization is established in those works.  The result needed here is its inefficient
conditional mixed-state extension within the controlled Hermitian diffusive model:
noncommuting control produces an exact nonnegative correction, which yields a
full-rank impurity-moment obstruction.

\emph{Determinant-root theorem.---}
For every full-rank initial state, every $\eta\in[0,1]$, every finite action $A$,
and every predictable policy in the control class of the Letter,
\begin{equation}
\E[R_A]\geq R_0e^{-2\eta V_LA}.
\label{eq:sm-detroot-bound}
\end{equation}
At $\eta=1$ equality holds for every policy.  At $0\leq\eta<1$, equality holds
whenever $[\rho_a,L_a]=0$ for $da\otimes d\mathbb P$-almost every
$(a,\omega)$; in particular it is attained by a fixed QND protocol initialized with
$[\rho_0,L]=0$, which therefore remains commuting.  The inequality is equally valid in the presence of predictable
finite-variation Hamiltonian control.

To prove the theorem, work in action time and omit the Hamiltonian term initially.
For traceless $L$, set
\begin{equation}
m=\Tr(\rho L),\qquad
G=\sqrt\eta\,(L\rho+\rho L-2m\rho),\qquad
\mathcal Q=\Tr(\rho^{-1}L\rho L).
\end{equation}
Thus $d\rho=\mathcal D[L]\rho\,da+G\,dW_a$.  Matrix It\^o calculus gives
\begin{equation}
d\ln\det\rho
=\Tr(\rho^{-1}d\rho)
-\frac12\Tr(\rho^{-1}G\rho^{-1}G)\,da.
\label{eq:sm-logdet-ito}
\end{equation}
The three required traces are
\begin{align}
\Tr[\rho^{-1}\mathcal D[L]\rho]&=\mathcal Q-\Tr L^2,
\label{eq:sm-logdet-drift1}\\
\Tr(\rho^{-1}G)&=-2d\sqrt\eta\,m,
\label{eq:sm-logdet-noise}\\
\Tr[(\rho^{-1}G)^2]
&=\eta\,[2\Tr L^2+2\mathcal Q+4dm^2].
\label{eq:sm-logdet-quad}
\end{align}
For Eq.~(\ref{eq:sm-logdet-quad}), put $C=\rho^{-1}L\rho$ and use
$\Tr C^2=\Tr L^2$, $\Tr(CL)=\mathcal Q$, and $\Tr C=\Tr L=0$.  Substitution into
Eq.~(\ref{eq:sm-logdet-ito}) yields
\begin{equation}
d\ln\det\rho
=\big[(1-\eta)\mathcal Q-(1+\eta)\Tr L^2-2\eta dm^2\big]da
-2d\sqrt\eta\,m\,dW_a.
\label{eq:sm-logdet}
\end{equation}
Applying scalar It\^o calculus to $R=(\det\rho)^{1/d}$ cancels the $m^2$ term:
\begin{align}
\frac{dR}{R}
&=\left[\frac{1-\eta}{d}\mathcal Q-(1+\eta)V_L\right]da
-2\sqrt\eta\,m\,dW_a\notag\\
&=\left[-2\eta V_L+\frac{1-\eta}{d}K(\rho,L)\right]da
-2\sqrt\eta\,m\,dW_a,
\label{eq:sm-detroot-sde}\\
K(\rho,L)&:=\mathcal Q-\Tr L^2.
\end{align}
In an eigenbasis $\rho=\operatorname{diag}(p_1,\ldots,p_d)$,
\begin{equation}
K(\rho,L)
=\sum_{i<j}\left(\frac{p_i}{p_j}+\frac{p_j}{p_i}-2\right)|L_{ij}|^2
=\sum_{i<j}\frac{(p_i-p_j)^2}{p_ip_j}|L_{ij}|^2\geq0.
\label{eq:sm-Kpositive}
\end{equation}
Moreover $K=0$ if and only if $[\rho,L]=0$.  Restoring a nonzero $\mu$ leaves
$K$ and $V_L$ invariant and changes only the noise coefficient in
Eq.~(\ref{eq:sm-detroot-sde}) to $2\sqrt\eta(\mu-\Tr\rho L)$.
The Hamiltonian contribution is zero because
$\Tr\{\rho^{-1}[-iH,\rho]\}=0$; finite unitary kicks preserve the determinant
exactly.

\begin{figure}[t]
\centering
\includegraphics[width=0.92\textwidth]{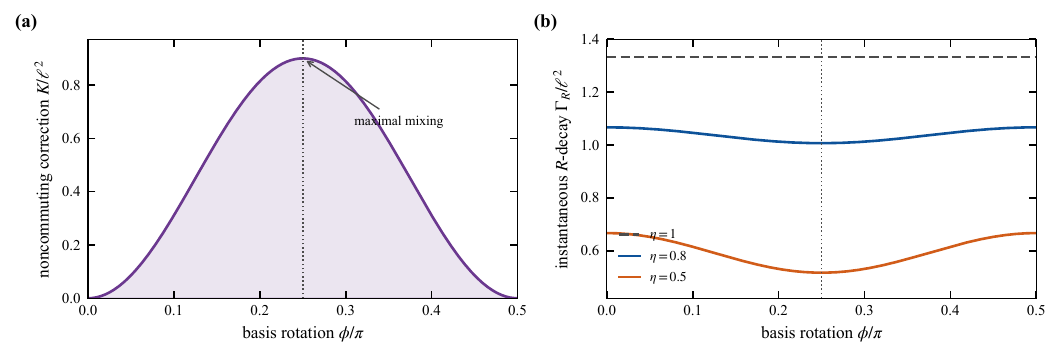}
\caption{Exact noncommuting correction for the full-rank state
$(p_-,p_0,p_+)=(0.50,0.30,0.20)$ and
$L_\phi=U_{-+}(\phi)\,\ell\,\mathrm{diag}(-1,0,1)U_{-+}^\dagger(\phi)$.
(a) $K/\ell^2=0.9\sin^2(2\phi)$ vanishes precisely at the commuting endpoints
of this rotation and is maximal at $\phi=\pi/4$.
(b) The conditional instantaneous determinant-root decay rate
$\Gamma_R/\ell^2=4\eta/3-(1-\eta)K/(3\ell^2)$ (with $dR/R=-\Gamma_R\,dA$ in the mean)
is orientation independent at
$\eta=1$ but is reduced by noncommuting orientations at $\eta<1$.  This panel
shows an instantaneous determinant-root law, not a fitted long-action exponent.}
\label{fig:sm-determinant}
\end{figure}

The correction has a direct physical consequence.  Figure~\ref{fig:sm-determinant} evaluates it
along a qutrit basis rotation and shows directly that loss converts noncommutativity
into a smaller instantaneous determinant-root decay rate.

We next justify boundary avoidance and removal of the localization.  Write
\begin{equation*}
\widehat L_a=L_a-\mu\openone,\qquad
\widehat m_a=\Tr(\rho_a\widehat L_a),\qquad
c_L=\|\widehat L_a\|_\infty .
\end{equation*}
The number $c_L$ is policy independent because the spectral multiset of $L_a$ is
fixed.  For
\begin{equation*}
\tau_n=\inf\{a\geq0:\lambda_{\min}(\rho_a)\leq n^{-1}\},
\qquad n>\lambda_{\min}(\rho_0)^{-1},
\end{equation*}
Equation~(\ref{eq:sm-logdet}), with
$\mathcal Q_a=dV_L+K_a$, has drift
$(1-\eta)K_a-2\eta dV_L-2\eta d\widehat m_a^2$.
Discarding $K_a\geq0$ up to $\tau_n$ gives, for $t\leq A$,
\begin{equation}
\ln\det\rho_{t\wedge\tau_n}
\geq\ln\det\rho_0-C_LA-\sup_{u\leq A}|\mathcal N_u|,
\qquad C_L=2\eta d(V_L+c_L^2),
\label{eq:sm-logdet-common-lower}
\end{equation}
where the same unstopped martingale is used for every $n$:
\begin{equation*}
\mathcal N_t=-2d\sqrt\eta\int_0^t\widehat m_a\,dW_a,\qquad
\langle\mathcal N\rangle_t\leq4d^2\eta c_L^2t.
\end{equation*}
Thus $\sup_{u\leq A}|\mathcal N_u|<\infty$ almost surely.  If rank were lost by
action $A$, continuity of the eigenvalues (unitary kicks leave them unchanged)
would imply $\tau_n\leq A$ for all sufficiently large $n$.  But then
$\det\rho_{\tau_n}\leq n^{-1}$, so the left side of
Eq.~(\ref{eq:sm-logdet-common-lower}) would be at most $-\ln n$, contradicting
its common finite lower bound.  Since $A<\infty$ was arbitrary, a full-rank
initial state remains full rank at every finite action almost surely.

Now set
\begin{equation*}
Y_a=e^{2\eta V_La}R_a,\qquad
\sigma_a=2\sqrt\eta\,(\mu-\Tr\rho_aL_a).
\end{equation*}
Equation~(\ref{eq:sm-detroot-sde}) becomes
\begin{equation}
dY_a=\frac{1-\eta}{d}Y_aK_a\,da+Y_a\sigma_a\,dW_a.
\label{eq:sm-Y-submartingale}
\end{equation}
On $[0,A]$, $Y_a\leq e^{2\eta V_LA}/d$ and
$|\sigma_a|\leq2\sqrt\eta c_L$, so the stochastic integral is a
square-integrable martingale.  Eigenvalue continuity and full-rank preservation give,
pathwise, $\delta_A:=\inf_{0\leq a\leq A}\lambda_{\min}(\rho_a)>0$.  Since
$0\leq K_a\leq\mathcal Q_a\leq\delta_A^{-1}c_L^2$, it follows explicitly that
$\int_0^A Y_aK_a\,da<\infty$ almost surely.  With
\begin{equation*}
\zeta_k=\inf\left\{t\geq0:\int_0^tY_aK_a\,da\geq k\right\},
\end{equation*}
optional stopping yields
\begin{equation}
\E Y_{A\wedge\zeta_k}
=Y_0+\frac{1-\eta}{d}
\E\!\int_0^{A\wedge\zeta_k}Y_aK_a\,da\geq Y_0.
\end{equation}
As $k\to\infty$, $A\wedge\zeta_k\to A$ almost surely, and the deterministic
bound on $Y$ permits dominated convergence.  Hence $\E Y_A\geq Y_0$, proving
Eq.~(\ref{eq:sm-detroot-bound}).  If $\eta=1$, or if $K_a=0$ for
$da\otimes d\mathbb P$-almost every $(a,\omega)$, the drift vanishes and the same
square-integrability bound makes $Y$ a true martingale, proving the stated
equalities.

\emph{Impurity-moment corollary.---}
Let $\alpha_d=(d-1)/d$, let $p=\max_i p_i$, and put
$\varepsilon=1-p$.  Since $p\geq1/d$,
\begin{equation}
S=2\sum_{i<j}p_ip_j\geq2p\varepsilon\geq\frac{2\varepsilon}{d},
\qquad
\det\rho\leq\left(\frac{\varepsilon}{d-1}\right)^{d-1},
\end{equation}
where the determinant bound is arithmetic--geometric mean on the other $d-1$
eigenvalues.  Consequently
\begin{equation}
S^{\alpha_d}\geq c_dR,
\qquad
c_d=\left[\frac{2(d-1)}d\right]^{\alpha_d}.
\label{eq:sm-SR}
\end{equation}
Because $0\leq S<1$, for $0<\theta\leq\alpha_d$,
Eqs.~(\ref{eq:sm-detroot-bound}) and (\ref{eq:sm-SR}) give
\begin{equation}
\E[S_A^\theta]\geq c_dR_0e^{-2\eta V_LA}.
\label{eq:sm-detmoment-low}
\end{equation}
For $\theta>\alpha_d$, set $q=\theta/\alpha_d>1$, raise
Eq.~(\ref{eq:sm-SR}) to the $q$th power, and use Jensen's inequality:
\begin{equation}
\E[S_A^\theta]
\geq c_d^q\E[R_A^q]
\geq(c_d\E R_A)^q
\geq(c_dR_0)^q e^{-2q\eta V_LA}.
\label{eq:sm-detmoment-high}
\end{equation}
Thus every fixed predictable policy from every full-rank initial state obeys
\begin{equation}
\limsup_{A\to\infty}-\frac1A\ln\E[S_A^\theta]
\leq2\eta V_L\max\left\{1,\frac{d\theta}{d-1}\right\}.
\label{eq:sm-detceiling}
\end{equation}
For $0<\eta<1$, this is to be combined by taking the minimum with the uniform
ceiling of Eq.~(\ref{eq:sm-alld-piecewise}).

For a nontrivial observable, the elementary range-variance inequality gives
$0<V_L\leq\ell^2$.  Equality occurs only when the empirical spectral measure puts equal mass at the two endpoints:
$d$ is even and $L$ is dichotomic with multiplicity $d/2$ at each endpoint.
Every generic nondegenerate spectrum in $d>2$ therefore has $V_L<\ell^2$ and a
strict gap from the rank-two frozen constant at least throughout
\begin{equation}
\frac12\leq\theta<\alpha_d\frac{\ell^2}{V_L}.
\label{eq:sm-detgapinterval}
\end{equation}
For odd $d$ one has the stronger maximum
$V_L\leq(1-d^{-2})\ell^2$: after scaling the eigenvalues into $[-1,1]$, variance
is convex in each entry and is maximized at a vertex, where the two endpoint
multiplicities must differ by at least one.  In the canonical qutrit case
$\operatorname{spec}L=(-\ell,0,+\ell)$, $V_L=2\ell^2/3$, and
Eq.~(\ref{eq:sm-detceiling}) becomes
\begin{equation}
\Lambda_\pi(\theta,\eta)\leq
\begin{cases}
4\eta\ell^2/3,&0<\theta\leq2/3,\\[2pt]
2\theta\eta\ell^2,&\theta>2/3.
\end{cases}
\label{eq:sm-qutritgap}
\end{equation}
For $0<\eta<1$, comparison with Eq.~(\ref{eq:sm-alld-piecewise}) excludes the
rank-two uniform ceiling throughout the larger interval
\begin{equation}
\frac{1-1/\sqrt3}{2}<\theta<1,
\end{equation}
under arbitrary predictable noncommuting control.  For the sparse family
$L_d=\operatorname{diag}(-\ell,0^{d-2},+\ell)$, one has
$V_L=2\ell^2/d$ and the same comparison gives
$[1-\sqrt{1-2/d}]/2<\theta<(d-1)/2$.  This interval collapses to the pivot only
at $d=2$.  The determinant lower bound does not constrain rank-deficient faces,
exactly as required by the embedded sharpness construction.  For balanced
dichotomic spectra or outside the interval supplied by the comparison, the
determinant theorem gives a necessary condition but does not settle attainability.

\begin{figure}[!t]
\centering
\includegraphics[width=0.92\textwidth]{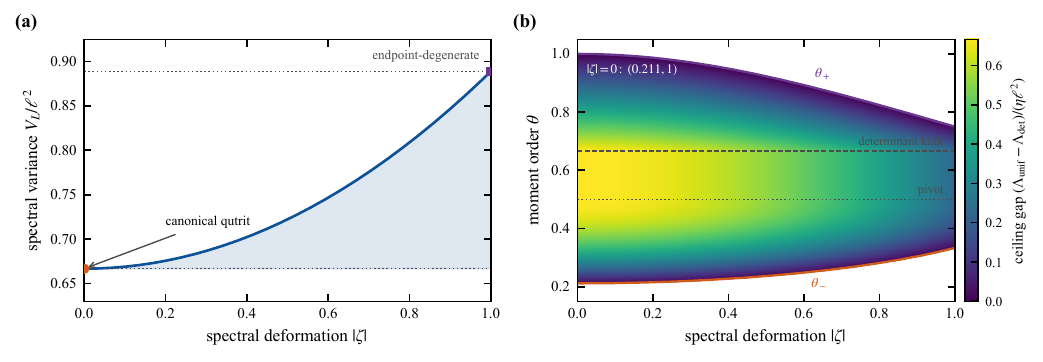}
\caption{Full-rank qutrit obstruction across
$L_\zeta=\ell\,\mathrm{diag}(-1,\zeta,1)$ for $0<\eta<1$.
(a) The normalized spectral variance $v_\zeta=2(3+\zeta^2)/9$; $\zeta=0$ is the
canonical equally spaced spectrum and $|\zeta|=1$ is endpoint-degenerate.
(b) The positive-part gap between the state-space-uniform ceiling and the
determinant ceiling, in units of $\eta\ell^2$.  It is positive precisely between
$\theta_-(\zeta)$ and $\theta_+(\zeta)$ in
Eq.~(\ref{eq:sm-qutrit-family-band}).  The colored band rigorously excludes
attainment of the rank-two uniform ceiling from a fixed full-rank initial state
under arbitrary predictable control; it delimits the obstruction interval.}
\label{fig:sm-qutrit-family}
\end{figure}

The dependence on the entire qutrit spectral family can be resolved analytically.
After removing an irrelevant scalar shift and fixing the half-spread, write
\begin{equation}
L_\zeta=\ell\,\operatorname{diag}(-1,\zeta,1),\qquad
-1\leq\zeta\leq1,\qquad
v_\zeta:=\frac{V_{L_\zeta}}{\ell^2}=\frac{2(3+\zeta^2)}9.
\end{equation}
For $0<\eta<1$, exact comparison of the determinant and uniform ceilings gives the strict
full-rank exclusion band
\begin{equation}
\theta_-(\zeta)<\theta<\theta_+(\zeta),\qquad
\theta_-:=\frac{1-\sqrt{1-v_\zeta}}2,\qquad
\theta_+:=\frac{2}{3v_\zeta}.
\label{eq:sm-qutrit-family-band}
\end{equation}
It ranges from $((1-1/\sqrt3)/2,1)$ at $\zeta=0$ to $(1/3,3/4)$ at
$|\zeta|=1$ and never closes.  Figure~\ref{fig:sm-qutrit-family} shows both the
spectral-variance deformation and the positive gap between the two ceilings.

\section{The qubit half-moment pivot law}
\label{sm:sqrt}

The Letter reduces the $\sqrt S$ drift to its aligned-endpoint value; we
establish that reduction in full here. Let $r(x,S)\equiv g(x,S)/S$. It\^o's formula applied to $\sqrt S$ with
Eq.~(\ref{eq:sm-sde}) gives
\begin{equation}
-\frac{\mathcal{A}\sqrt S}{\sqrt S}
= M\Big[\,r(x,S) + 2\eta\,x\beta\,\Big],
\label{eq:sm-sqrtgen}
\end{equation}
the second term being the It\^o correction
$\tfrac12\cdot\tfrac{1}{4}S^{-2}\cdot(4\sqrt{\eta M x\beta}S)^{2}/M$. The bracket is
affine in $x$ with slope
\begin{equation}
\partial_x\big[r+2\eta x\beta\big]
=(1-\eta)\,\frac{(1-S)-2S^{2}}{S}-(1+\eta)\beta+2\eta\beta
=\frac{(1-\eta)(1-2S)}{S}\;\ge\;0 ,
\end{equation}
so its maximum over the basis sits at the aligned endpoint, where
$g(\ell^{2},S)=4\eta\ell^{2}S^{2}$ and
\begin{equation}
\Big[\,r+2\eta x\beta\,\Big]_{x=\ell^{2}}
= 4\eta\ell^{2}S + 2\eta\ell^{2}(1-2S) \;\equiv\; 2\eta\ell^{2},
\label{eq:sm-pivot}
\end{equation}
a constant, independent of the state. This proves the qubit pivot bound: for every adapted protocol
$-\mathcal{A}\sqrt S\le 2\eta\ell^{2}M\sqrt S$, with equality if and only if
$x=\ell^{2}$ when $0\le\eta<1$ and $0<S<\tfrac12$.  At the maximally mixed
state $S=\tfrac12$ the slope also vanishes, so every basis is instantaneously an
equality case; at $\eta=1$ equality holds for every basis at every mixed state.
For a deterministic action $A_T=A$, taking expectations and applying
Gr\"onwall's inequality in the action clock $dA=M\,dt$ yields the finite-action
half-moment bound in the Letter.
More generally, if $M_t$ is adapted and $A_T=\int_0^T M_tdt\le\bar A$ almost
surely, then $e^{2\eta\ell^2A_t}\sqrt{S_t}$ is a local submartingale
bounded above by $e^{2\eta\ell^2\bar A}/\sqrt2$, hence a true submartingale to
which optional stopping applies, and
$\E\sqrt{S_T}\ge\sqrt{S_0}e^{-2\eta\ell^2\bar A}$.  No deterministic
exponential may be pulled outside the expectation for an unrestricted random
action. Because Eq.~(\ref{eq:sm-pivot}) holds pointwise, the
aligned protocol attains the half-moment bound with equality at every instant; at $\eta=1$,
$e^{2\ell^{2}A_t}\sqrt{\det\rho_t}$ is a local martingale for \emph{every}
protocol (and therefore has constant expectation under a deterministic action
horizon or the bounded-action localization used above).
(Regularity: from a mixed initial state the process never reaches $S=0$
at any finite action under any admissible protocol, by the full-rank preservation
argument of Sec.~\ref{sm:detlaw} applied to $\det\rho=S/2$, so It\^o's formula
applies without boundary terms.)

\section{Dimension-independence of the moment envelope}
\label{sm:ddimmoment}

The general-dimension frame of Sec.~\ref{sm:ddim} carries the entire moment
envelope of the Letter into arbitrary finite dimension $d$, at fixed spectral
half-spread $\ell$ of the measured observable. The master inequality $C_{\sqrt S}$
[Eq.~(\ref{sm:csqrt})] and the pairing bound of Lemma~A [Sec.~\ref{sm:lemA}] are the
only two inputs; the qubit is recovered as the two-level aligned face on which both
hold with equality.

\emph{Reduction theorem.---}Assume $C_{\sqrt S}$ and Lemma~A. Then in every
dimension, for every adapted protocol and every fixed moment order $\theta>0$,
\begin{equation}
-\frac{\mathcal{A}S^{\theta}}{S^{\theta}}\;\le\;c(\theta)\,\eta\ell^{2}M,
\qquad
c(\theta)=\begin{cases}8\theta(1-\theta), & \theta\le\tfrac12,\\[2pt]
4\theta,&\theta\ge\tfrac12,\end{cases}
\label{sm:reduction}
\end{equation}
Consequently,
$\E[S_{T}^{\theta}]\ge S_{0}^{\theta}e^{-c(\theta)\eta\ell^{2}A}$ when the
terminal action $A_T=A$ is deterministic.  The same conclusion with $A$ replaced
by $\bar A$ holds for an adapted strength satisfying $A_T\le\bar A$ almost
surely.  This is a dimension-independent lower envelope; sharpness of every
branch in a full-rank $d$-level problem is not asserted; Sec.~\ref{sm:detlaw}
gives a spectral condition under which it is impossible near the pivot.

\emph{Proof} (three lines of It\^o algebra). The impurity of Sec.~\ref{sm:ddim}
obeys $dS=2M[(1-\eta)T_{1}-(1+\eta)T_{2}]\,dt-4\sqrt{\eta M}\,T_{3}\,dW$, with squared
diffusion $16\eta M\,T_{3}^{2}$. It\^o's formula for $S^{\theta}$ gives
\begin{equation}
-\frac{\mathcal{A}S^{\theta}}{S^{\theta}}
=2\theta M\,\frac{(1+\eta)T_{2}-(1-\eta)T_{1}}{S}
+8\theta(1-\theta)\,\eta M\,\frac{T_{3}^{2}}{S^{2}} .
\end{equation}
Bounding the first ratio with $C_{\sqrt S}$, i.e.
$(1+\eta)T_{2}-(1-\eta)T_{1}\le 2\eta\ell^{2}S-2\eta T_{3}^{2}/S$, collapses this to
\begin{equation}
-\frac{\mathcal{A}S^{\theta}}{S^{\theta}}
\;\le\;4\theta\eta\ell^{2}M+4\theta(1-2\theta)\,\eta M\,\frac{T_{3}^{2}}{S^{2}} .
\end{equation}
For $\theta\le\tfrac12$ the residual term is nonnegative and Lemma~A
($T_{3}^{2}\le\ell^{2}S^{2}$) caps it, giving $8\theta(1-\theta)\eta\ell^{2}M$; for
$\theta\ge\tfrac12$ it is nonpositive and is discarded, giving $4\theta\eta\ell^{2}M$.
The exponential transform
$e^{c(\theta)\eta\ell^{2}A_t}S_t^\theta$ and localization then give the two
action statements following Eq.~(\ref{sm:reduction}). $\square$

The pivotal case $\theta=\tfrac12$ is the dimension-independence of the
$\sqrt S$ speed limit, $\E[\sqrt{S_{T}}]\ge\sqrt{S_{0}}\,e^{-2\eta\ell^{2}A}$ in all
$d$, the general half-moment statement in the Letter. For $d>2$ this is a law on $\E[\sqrt S]$,
not on $\E[\sqrt{\det\rho}\,]$; the two coincide only on the qubit face.

\emph{First-passage corollary} $(\theta\to0)$. Dividing the
$\theta\le\tfrac12$ bound by $\theta$ and letting $\theta\to0$ turns
Eq.~(\ref{sm:reduction}) into the pointwise logarithmic bound
\begin{equation}
-\mathcal{A}\ln S\;\le\;8\eta\ell^{2}M \qquad(\text{all }d),
\end{equation}
the general-$d$ logarithmic speed limit used in Sec.~\ref{sm:fp}. The optional-stopping
argument there is itself dimension-independent, so it carries the first-passage bound
$\E[A(T_{\epsilon})]\ge\ln(S_{0}/\epsilon)/(8\eta\ell^{2})$ into every dimension; the
full stopping construction is written out as the $\theta\to0$ extension of
Sec.~\ref{sm:fp}.

\emph{Sharp embedded faces.---}Let $|+\rangle,|-\rangle$ be eigenvectors of $L$ with
the extreme eigenvalues and choose the diagonal mixture
$\rho=p|+\rangle\langle+|+(1-p)|-\rangle\langle-|$ with $0<p<1$.  After an irrelevant identity
shift, $L=\ell\sigma_z$ on this block, so $[\rho,L]=0$ and the configuration is
aligned (QND); the centered operator remains
$\tilde L=L-\Tr(\rho L)\openone$.  The diffusive dynamics of Sec.~\ref{sm:exact} keeps
$\rho$ within the block, and $S=1-\Tr\rho^{2}=2p(1-p)$ is the qubit
impurity. The channel is therefore \emph{exactly} the qubit aligned channel, on which
$C_{\sqrt S}$ and $C_{T2}'$ [Sec.~\ref{sm:ct2}] hold with equality, and its moment
spectrum is the qubit QND spectrum of Sec.~\ref{sm:exact},
\begin{equation}
\Lambda_{\rm QND}(\theta)=
\begin{cases}8\theta(1-\theta)\,\eta\ell^{2}, & \theta\le\tfrac12,\\[2pt]
2\eta\ell^{2}, & \theta\ge\tfrac12 ,\end{cases}
\qquad(0<\eta\le1).
\label{sm:embed}
\end{equation}
Hence the envelope constant $c(\theta)$ of Eq.~(\ref{sm:reduction}) is attained
in exponent on this embedded face for $\theta\le\tfrac12$, while for $\theta\ge\tfrac12$ the
attained QND exponent \emph{freezes} at its $\theta=\tfrac12$ value $2\eta\ell^{2}$
(the dimension-independent ceiling of Sec.~\ref{sm:alldfreeze} for $\eta<1$,
below the finite-action envelope $4\theta\eta\ell^{2}$).
At $\eta=1$ the two-level-embedded unbiased channel instead attains the Jacobs line
$4\theta\ell^{2}=c(\theta)\ell^{2}$ for $\theta\ge\tfrac12$.
For $0<\eta<1$, Eq.~(\ref{sm:embed}) consequently attains every branch of the
piecewise uniform ceiling in Sec.~\ref{sm:alldfreeze}.  This is sharpness over the
allowed state space on a rank-two invariant face, not generic full-rank
attainability.  Indeed, Sec.~\ref{sm:detlaw} proves a strict interior gap for every
non-balanced spectrum over a nonempty moment interval.  In particular, a $d=3$
realization with zero population in the intermediate eigenspace is exactly this
embedded qubit face, not evidence for generic full-rank qutrit saturation.

\section{Dimension-independent inefficient moment ceiling}
\label{sm:alldfreeze}

We now improve the high-moment branch of the finite-action envelope at long action.
The result is uniform in dimension and protocol, but it is an asymptotic moment
bound, not a fixed-threshold survival theorem.

\emph{Theorem (all-dimensional inefficient ceiling).---}
Fix a finite dimension $d$, a mixed initial state $S_0>0$, an efficiency
$0<\eta<1$, and an observable of spectral half-spread $\ell$.  For every fixed
$\theta>0$ and every fixed predictable policy in the control class of the Letter,
\begin{equation}
\limsup_{A\to\infty}-\frac{1}{A}\ln\E_{\pi}[S_A^\theta]
\leq 2\eta\ell^2 .
\label{eq:sm-alld-ceiling}
\end{equation}
In particular, this controls the $\liminf$ exponent $\Lambda_\pi$ defined in the
Letter.  Combining Eq.~(\ref{eq:sm-alld-ceiling}) with the stronger low-order
finite-action envelope of Eq.~(\ref{sm:reduction}) gives
\begin{equation}
\Lambda_\pi(\theta,\eta)\leq
\begin{cases}
8\theta(1-\theta)\eta\ell^2,&0<\theta\leq\tfrac12,\\[2pt]
2\eta\ell^2,&\theta\geq\tfrac12,
\end{cases}
\qquad 0<\eta<1 .
\label{eq:sm-alld-piecewise}
\end{equation}
The constants in Eq.~(\ref{eq:sm-alld-piecewise}) cannot be lowered uniformly over
the allowed state space: the extremal rank-two QND face of
Eq.~(\ref{sm:embed}) attains both branches.  This uniform ceiling is not generally
attained from a full-rank initial state: the determinant theorem of
Sec.~\ref{sm:detlaw} rules it out for generic spectra near the pivot.
If $\ell=0$, the measurement is scalar up to an irrelevant identity shift and the
claim is immediate; hence assume $\ell>0$ below.

The proof has two ingredients.  The first is a rigidity estimate showing that near
the pure-state boundary any drift capable of exponential purification must carry
nearly maximal innovation noise.

\emph{Near-pure rigidity lemma.---}
With the trace frame of Eq.~(\ref{sm:ddim-frame}), put
$\Delta=T_1-T_2\geq0$.  Whenever $S\leq\tfrac14$,
\begin{equation}
T_2\leq 12\ell^2S^2+8S\Delta .
\label{eq:sm-nearpure}
\end{equation}
To prove this, diagonalize $\rho=\operatorname{diag}(p_1,\ldots,p_d)$ with
$p_1=\max_i p_i$ and write $\varepsilon=1-p_1$.  Since
$\sum_i p_i^2\leq p_1$, one has $\varepsilon\leq S$; hence $S\leq1/4$ implies
$p_1\geq3/4$.  Shift $L$ by the midpoint of its spectral interval so that
$\|L\|\leq\ell$, and put $m_i=\widetilde L_{ii}$.  Centering gives
\begin{equation}
m_1=\sum_{j>1}p_j(L_{11}-L_{jj}),
\end{equation}
so $|m_1|\leq2\ell\varepsilon$ and, for $j>1$,
$|m_j|\leq|L_{jj}-L_{11}|+|m_1|\leq(5/2)\ell$.  In this basis,
\begin{align}
T_2&=\sum_i p_i^2m_i^2
 +2\sum_{i<j}p_ip_j|L_{ij}|^2,
\label{eq:sm-t2split}\\
\Delta&=\sum_{i<j}(p_i-p_j)^2|L_{ij}|^2 .
\label{eq:sm-deltasplit}
\end{align}
The diagonal part of Eq.~(\ref{eq:sm-t2split}) and the pairs with $i,j>1$ obey
\begin{equation}
\sum_i p_i^2m_i^2+2\sum_{2\leq i<j}p_ip_j|L_{ij}|^2
\leq\left(4+\frac{25}{4}+1\right)\ell^2\varepsilon^2
\leq12\ell^2S^2.
\end{equation}
For every dominant pair $(1,j)$,
$2p_1p_j\leq2\varepsilon\leq8\varepsilon(p_1-p_j)^2$, because
$p_1-p_j\geq1/2$.  Summing and using $\varepsilon\leq S$ proves
Eq.~(\ref{eq:sm-nearpure}).

For the second ingredient, work directly in the action clock and define
\begin{equation}
B=(1+\eta)T_2-(1-\eta)T_1
=2\eta T_2-(1-\eta)\Delta .
\label{eq:sm-Bdef}
\end{equation}
Equations~(\ref{eq:sm-nearpure}) and (\ref{eq:sm-Bdef}) give
\begin{equation}
B\leq24\eta\ell^2S^2+\big[16\eta S-(1-\eta)\big]\Delta .
\end{equation}
Thus, with
\begin{equation}
s_\eta=\min\left\{\frac14,\frac{1-\eta}{16\eta}\right\},
\qquad S\leq s_\eta,
\end{equation}
one has the protocol-independent local estimate
\begin{equation}
B\leq24\eta\ell^2S^2.
\label{eq:sm-Blocal}
\end{equation}

Set
\begin{equation}
X=\frac12\ln S,
\qquad \nu=-2\sqrt\eta\,\frac{T_3}{S}.
\end{equation}
It\^o's formula and Eq.~(\ref{sm:ddim-sde}) yield exactly
\begin{equation}
dX=-\left(\frac{B}{S}+\nu^2\right)da+\nu\,dW_a,
\qquad \nu^2\leq4\eta\ell^2=:r_{\max},
\label{eq:sm-logS}
\end{equation}
where the last inequality is Lemma~A.  Define $\nu=0$ on $\{S=0\}$; the same bound
makes this a bounded predictable extension, and put
$N_t=\int_0^t\nu_a\,dW_a$.  All logarithmic calculations may first be stopped at
$\tau_n=\inf\{a:S_a\leq1/n\}$.  Fix $A<\infty$ and $s<s_\eta$.  For all
sufficiently large $n$, if $\tau_n\leq A$, integrate
Eq.~(\ref{eq:sm-logS}) after the last crossing of $S=s$
(or from zero if the path starts below $s$ and never reaches it).  With
$c_0=\min\{(1/2)\ln S_0,(1/2)\ln s\}$ this gives the common, $n$-independent bound
\begin{equation}
X_{\tau_n}\geq c_0-(24\eta\ell^2s+r_{\max})A
-2\sup_{u\leq A}|N_u|.
\end{equation}
The right-hand side is finite because
$\langle N\rangle_A\leq r_{\max}A$.  It cannot hold for arbitrarily large $n$,
since $X_{\tau_n}=-(1/2)\ln n$.  Thus the stopping is removable and $S_a>0$ at
every finite action for every mixed initial state, including rank-deficient ones.

Fix $0<s<s_\eta$ and a deterministic action horizon $A$.  Set
$\chi_a=\mathbf1_{\{S_{a-}\leq s\}}$; continuity makes it equal to
$\mathbf1_{\{S_a\leq s\}}$ in all integrals, while the left-limit form is explicitly
predictable.  Define
on $[0,A]$
\begin{equation}
Z_t=\exp\left\{\int_0^t\chi_a\nu_a\,dW_a
-\frac12\int_0^t\chi_a\nu_a^2\,da\right\}.
\label{eq:sm-girsanov-density}
\end{equation}
The integrand is bounded, so Novikov's condition holds and $Z_t$ is a true
martingale.  Under the equivalent measure $d\mathbb Q_A=Z_A\,d\mathbb P$,
\begin{equation}
W_t^{\mathbb Q}=W_t-\int_0^t\chi_a\nu_a\,da
\end{equation}
is Brownian motion.  Equivalence preserves the filtration and therefore the
predictability of every record-dependent policy.  Equation~(\ref{eq:sm-logS}) becomes
\begin{equation}
dX=\left[-\frac{B}{S}-(1-\chi)\nu^2\right]da
+\nu\,dW_a^{\mathbb Q}.
\label{eq:sm-logS-Q}
\end{equation}
In the low-impurity region its drift is at least
$-\kappa_s$, where $\kappa_s=24\eta\ell^2s$.

Let
\begin{equation}
\mathcal M_t=\int_0^t\chi_a\nu_a\,dW_a^{\mathbb Q},
\qquad \mathcal V_t=\langle\mathcal M\rangle_t
=\int_0^t\chi_a\nu_a^2\,da
\leq r_{\max}t .
\end{equation}
Put $c=(1/2)\ln s$ and $c_0=\min\{X_0,c\}$.  If $X_A<c$, integrate
Eq.~(\ref{eq:sm-logS-Q}) after the last crossing of $c$; if that crossing does not
exist, the path started below $c$ and the same estimate begins at zero.  If
$X_A\geq c$, the resulting lower bound is automatic.  In all cases,
\begin{equation}
X_A\geq c_0-\kappa_sA-2\sup_{t\leq A}|\mathcal M_t|.
\label{eq:sm-last-excursion}
\end{equation}
For $\gamma>0$ set
$F_A=\{\sup_{t\leq A}|\mathcal M_t|\leq\gamma A\}$.  The exponential martingale maximal
inequality (equivalently, Dambis--Dubins--Schwarz followed by the Brownian maximal
bound) gives
\begin{equation}
\mathbb Q_A(F_A^c)\leq
2\exp\left[-\frac{\gamma^2A}{2r_{\max}}\right]
=2\exp\left[-\frac{\gamma^2A}{8\eta\ell^2}\right].
\label{eq:sm-Mmax}
\end{equation}
Writing Eq.~(\ref{eq:sm-girsanov-density}) in terms of $W^{\mathbb Q}$ gives
\begin{equation}
\left.\frac{d\mathbb P}{d\mathbb Q_A}\right|_{\mathcal F_A}
=Z_A^{-1}=\exp\left[-\mathcal M_A-\frac12\mathcal V_A\right].
\label{eq:sm-reverse-density}
\end{equation}
On $F_A$, Eqs.~(\ref{eq:sm-last-excursion})--(\ref{eq:sm-reverse-density}) imply,
for all sufficiently large $A$,
\begin{align}
\E_{\mathbb P}[S_A^\theta]
&\geq\left(1-2e^{-\gamma^2A/(8\eta\ell^2)}\right)
\min\{S_0,s\}^{\theta}\notag\\
&\quad\times
\exp\left[-\big\{2\eta\ell^2+48\theta\eta\ell^2s
+(1+4\theta)\gamma\big\}A\right].
\label{eq:sm-corridor-bound}
\end{align}
First take $A\to\infty$ with $s$ and $\gamma$ fixed, then let
$\gamma\downarrow0$ and $s\downarrow0$.  This proves
Eq.~(\ref{eq:sm-alld-ceiling}).  The measures $\mathbb Q_A$ need not be consistent
between different horizons; Eq.~(\ref{eq:sm-corridor-bound}) is a physical-measure
inequality at each fixed $A$.  The policy is fixed before the long-action limit and
is parametrized in the action clock; the equivalent physical-time statement uses
the inverse clock of the Letter and assumes unbounded accumulated action.  The
order $\theta$ is fixed, finite, and positive.  The mixed-state hypothesis
$S_0>0$ is essential; a pure initial state is not a purification problem.  At
$\eta=0$, Eq.~(\ref{sm:ddim-sde}) gives
$dS=2(T_1-T_2)\,da\geq0$, so the zero ceiling is trivial.  The restriction
$\eta<1$ in the nontrivial proof is essential because $s_\eta$ collapses at perfect
efficiency, where unbiased feedback can exceed the frozen branch.

\emph{Why the qubit fixed-threshold certificate does not extend directly.---}
The change of measure is what makes the argument work in higher dimension, where no scalar
verification function is available.  In $d=3$, take
\begin{equation}
\rho=\operatorname{diag}(1-2\varepsilon,\varepsilon,\varepsilon),
\qquad
L=\begin{pmatrix}
0&0&0\\
0&0&\ell\\
0&\ell&0
\end{pmatrix}.
\end{equation}
Then $T_3=0$ and $T_1=T_2=2\ell^2\varepsilon^2$, so
$B=4\eta\ell^2\varepsilon^2>0$: the impurity drifts toward the pure boundary with
zero instantaneous $S$-diffusion.  A nonnegative verification function depending
only on $S$, vanishing at a fixed lower threshold and increasing into the mixed
region, would therefore have a negative generator at that boundary and cannot obey
the qubit submartingale inequality.  Crucially this quiet purification direction is
only $O(S^2)$, so it has no positive exponential rate; Eq.~(\ref{eq:sm-Blocal}) is
exactly what allows the localized change-of-measure proof to retain the sharp
long-action ceiling.

\section{Qubit endpoint proof of the moment envelope}
\label{sm:moments}

For every fixed $\theta>0$, the same computation applied to $S^{\theta}$ gives
\begin{equation}
-\frac{\mathcal{A}S^{\theta}}{S^{\theta}}
= \theta M\Big[\,\frac{2g(x,S)}{S} + 8(1-\theta)\,\eta\,x\beta\,\Big],
\label{eq:sm-momgen}
\end{equation}
again affine in $x$. At the aligned endpoint the bracket equals
$8\eta\ell^{2}[(1-\theta)+S(2\theta-1)]$; at the unbiased endpoint it equals
$2\ell^{2}[(1+\eta)-(1-\eta)(1-S)/S]$.  Neither endpoint dominates the other
for every $(S,\theta,\eta)$: at $\eta=1$, $\theta=1$, and $S=\tfrac14$, for
example, the unbiased value is twice the aligned one.  We therefore bound the
endpoints separately.  The aligned endpoint gives
\begin{equation}
\begin{aligned}
&\sup_S 8\theta\eta\ell^2\big[(1-\theta)+S(2\theta-1)\big]
=c(\theta)\eta\ell^2,\\
&c(\theta)=\begin{cases}
8\theta(1-\theta), & \theta\le\tfrac12 \quad (\text{supremum at } S\to0),\\[2pt]
4\theta, & \theta\ge\tfrac12 \quad (\text{supremum at } S=\tfrac12),
\end{cases}
\end{aligned}
\end{equation}
whereas the unbiased expression is increasing in $S$ and hence
\begin{equation}
2\theta\ell^2\Big[(1+\eta)-(1-\eta)\frac{1-S}{S}\Big]
\le4\theta\eta\ell^2\le c(\theta)\eta\ell^2.
\end{equation}
For $\theta\ge\tfrac12$ the last inequality is an equality; for
$\theta\le\tfrac12$ it is equivalent to $1-\theta\ge\tfrac12$.  Affinity in
$x$ now proves the envelope, and the exponential-submartingale argument of
Sec.~\ref{sm:ddimmoment} proves the Letter's finite-action moment envelope under the deterministic-action or
almost-sure action-cap condition stated there. Note that for $\theta<\tfrac12$ the
supremum is approached only as $S\to0$: the bound is nevertheless exactly attained in
exponent, because the aligned channel spends its late time at small $S$
(Sec.~\ref{sm:exact}).

\begin{figure}[!t]
\centering
\includegraphics[width=0.92\textwidth]{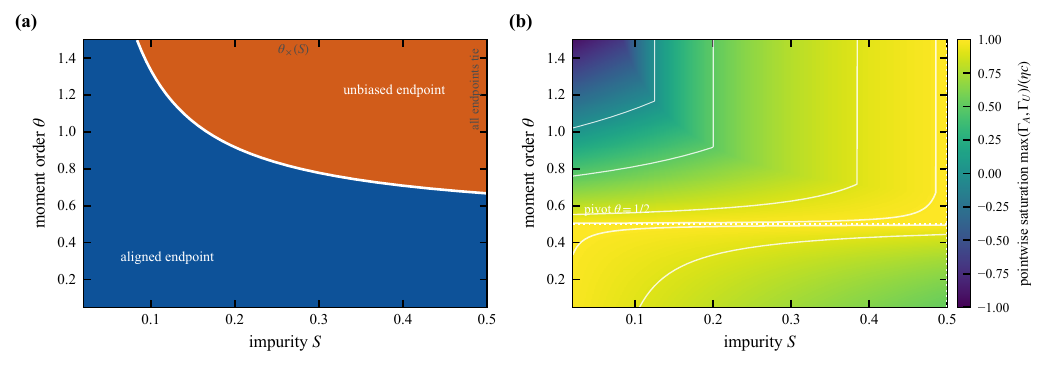}
\caption{Pointwise qubit moment-generator geometry at $\eta=3/4$.
(a) The endpoint with the larger instantaneous decay generator.  The white curve
is the exact switch line $\theta_\times(S)$; all orientations tie at the maximally
mixed boundary $S=1/2$.
(b) The larger endpoint generator divided by the global envelope
$\eta c(\theta)$.  Saturation occurs as $S\to0$ below the pivot, along the entire
line $\theta=1/2$, and at $S=1/2$ above it.  Negative values mean that even the
larger endpoint generator instantaneously increases the moment.  This is a
pointwise generator map, not a finite-horizon switching optimum or a long-action
rate diagram.}
\label{fig:sm-moment-geometry}
\end{figure}

The endpoint competition itself has a simple geometry that is hidden by taking the
two suprema separately.  Define the dimensionless instantaneous decay generators
$\Gamma_A$ and $\Gamma_U$ by
$\Gamma=-\mathcal A S^\theta/(M\ell^2S^\theta)$ at the aligned and unbiased
endpoints.  Their difference factorizes exactly as
\begin{equation}
\Gamma_A-\Gamma_U
=\frac{2\theta(2S-1)}{S}
\left(4\eta S\theta-2\eta S+\eta-1\right).
\label{eq:sm-endpoint-switch}
\end{equation}
Hence, for $0<\eta\le1$ and $0<S<1/2$, the aligned endpoint is faster below and the unbiased
endpoint is faster above
\begin{equation}
\theta_\times(S)=\frac12+\frac{1-\eta}{4\eta S};
\label{eq:sm-switch-line}
\end{equation}
at $S=1/2$ every orientation ties.  Figure~\ref{fig:sm-moment-geometry} also
shows where the pointwise generator saturates the global envelope used above.

\section{Exact solution of the plain-measurement channel at arbitrary efficiency}
\label{sm:exact}

Throughout this section $0<\eta\le1$, $\ell>0$, and the initial state
is mixed and aligned: $0<S_0\le\tfrac12$ and $[\rho_0,L]=0$, equivalently
$y_0\in\mathbb R$.  The pure boundary corresponds to $|y_0|=\infty$ and is excluded;
every rate below divides by $a=\eta\ell^2>0$.

At $x=\ell^{2}$ the state remains diagonal and Eq.~(\ref{eq:sm-sde}) reduces, in
the action clock $A=\int Mdt$, to the closed one-dimensional diffusion
$dz=2\ell\sqrt{\eta}\,(1-z^{2})\,dW_A$ for the
population gap $z=\pm\sqrt{\beta}$ ($S=(1-z^{2})/2$). The substitution
$y=\operatorname{arctanh} z$ maps it, by It\^o's formula, to
\begin{equation}
dy = \kappa\,\eta \tanh y \, dA + \sqrt{\kappa\,\eta}\; dW_A,
\qquad \kappa\equiv4\ell^{2},
\label{eq:sm-y}
\end{equation}
a Brownian motion with $\tanh$ drift and no boundary ($S=\tfrac12\operatorname{sech}^{2}y$).
The action generator $\mathcal{L}_A=\tfrac{\kappa\eta}{2}\partial_{y}^{2}+\kappa\eta
\tanh(y)\partial_y$ is the gradient form associated with the measure
$\cosh^{2}y\,dy$; the ground-state transform $f\mapsto \cosh(y)\,f$ maps it to a
Schr\"odinger operator whose potential,
$\tfrac{\kappa\eta}{2}(\tanh^{2}y+\operatorname{sech}^{2}y)=\tfrac{\kappa\eta}{2}$,
is exactly constant. Hence
\begin{equation}
\E_{y_0}\big[f(y_A)\big]
= \frac{e^{-\kappa\eta A/2}}{\cosh y_0}\;
\big(G_{\kappa\eta A}\ast \cosh\!\cdot\! f\big)(y_0),
\label{eq:sm-heat}
\end{equation}
with $G_v$ the centered Gaussian kernel of variance $v$: the channel is a free heat
flow in disguise, at every efficiency.  Equation~(\ref{eq:sm-heat}) is an equivalent
heat-kernel representation of the known exact QND transition law used in
Refs.~\cite{Li2013,Jiang2020}.  For a maximally mixed initial state and after matching
time and measurement-rate conventions, Eq.~(\ref{eq:sm-heat}) coincides with the exact
Onsager--Machlup path-integral propagator of Ref.~\cite{Poltronieri2026}.  The use made here
of this known solvable channel is to extract the full moment spectrum and provide a
matching protocol.  Three consequences used in the Letter are:
(i) for $f=\operatorname{sech}$, $\cosh\cdot f=1$ and
$\E[\sqrt{S_A}]=\sqrt{S_0}\,e^{-2\eta\ell^{2}A}$ \emph{exactly};
(ii) for $f=\operatorname{sech}^{2\theta}$, a Gaussian
integral of $\cosh^{1-2\theta}$ gives the QND moment spectrum,
$\Lambda_{\rm QND}(\theta)=\tfrac{\kappa\eta}{2}\{1-(1-2\theta)^{2}\}
=8\theta(1-\theta)\eta\ell^{2}$ for $\theta\le\tfrac12$, and the frozen value
$2\eta\ell^{2}$ for $\theta\ge\tfrac12$.  At the pivot
$\theta=\tfrac12$ the decay is the exact pure exponential in (i), with no
algebraic prefactor.  Only for $\theta>\tfrac12$ is
$\cosh^{1-2\theta}$ integrable, in which case the spectral edge supplies the
$A^{-1/2}$ prefactor; (iii) the process
never leaves finite $y$ in finite time, providing the regularity used in
Sec.~\ref{sm:fp}.

A further consequence, used in the Fig.~1 caption of the Letter, is the
fixed-threshold endpoint asymptotics of this exact channel.  For $0<\eta\le1$,
$\ell>0$, a mixed initial state ($0<S_0\le\tfrac12$), and a fixed threshold
$0<s<\tfrac12$, the event $\{S_A\geq s\}$ confines
$y_A$ to the bounded window $|y|\leq\operatorname{arcsech}\sqrt{2s}$, on which the
Gaussian factor of Eq.~\eqref{eq:sm-heat} is $\Theta(v^{-1/2})$ uniformly as
$A\to\infty$; hence
\begin{equation}
\Pr_{\rm QND}\big[S_A\geq s\big]
=e^{-2\eta\ell^{2}A+o(A)} .
\label{eq:sm-qnd-survival}
\end{equation}
A separate pathwise no-hit certificate, at the threshold-specific rate
$2\eta\ell^{2}(1+\omega^{2})$ for every adapted protocol, is constructed in
Sec.~\ref{sm:frozen}: under an action budget $A_T\le\bar A$ it lower-bounds
$\Pr[\tau_Y>T]$, an event contained in $\{S_T\geq s_Y\}$; its certified rate
approaches $2\eta\ell^{2}$ in the ordered limits $\bar A\to\infty$, then
$\omega\downarrow0$.

\emph{Large-deviation form and termination.---}
Equation~\eqref{eq:sm-heat} also gives the transition density
\begin{equation}
p_A(y\mid y_0)=
\frac{e^{-v/2}\cosh y}{\cosh y_0}\,G_v(y-y_0),
\qquad v=4\eta\ell^2A .
\label{eq:sm-ydensity}
\end{equation}
For $0<\eta\leq1$, put $a=\eta\ell^2$ and $q_A=y_A/A$.  Reading the exponential scale of
Eq.~\eqref{eq:sm-ydensity} at $y=qA$ gives the rate function
\begin{equation}
J(q)=2a+\frac{q^2}{8a}-|q|
=\frac{(|q|-4a)^2}{8a}.
\label{eq:sm-qrate}
\end{equation}
For $X_A=-A^{-1}\ln S_A$, the identity
$-\ln S_A=\ln2+2\ln\cosh y_A$ implies
$|AX_A-2|y_A||\leq\ln2$.  The contraction principle therefore gives
\begin{equation}
I(x)=\frac{(x-8a)^2}{32a},\qquad x\geq0 ,
\label{eq:sm-xrate}
\end{equation}
and the Laplace principle yields
\begin{equation}
\begin{split}
\Lambda_{\rm QND}(\theta,\eta)
&=\inf_{x\geq0}\{I(x)+\theta x\},\\
x_\theta&=\max\{8a(1-2\theta),0\}.
\end{split}
\label{eq:sm-legendre}
\end{equation}
The minimizing saddle reaches the physical boundary $x=0$ at
$\theta=\tfrac12$ and remains there above it.  This boundary-saddle
termination is exactly the piecewise spectrum in Eq.~\eqref{sm:embed}; the
all-dimensional ceiling of Sec.~\ref{sm:alldfreeze} makes it optimal over all
adapted qubit protocols when $0<\eta<1$, while the separate theorem in
Sec.~\ref{sm:frozen} strengthens that moment statement to fixed-threshold
qubit survival.

\emph{Critical moment-order layer.---}
The change of saddle at $\theta=\tfrac12$ has a uniform boundary-layer
resolution.  Fix $0<S_0\leq\tfrac12$ (hence finite $y_0$), let
$v=4\eta\ell^2A\to\infty$, and set
\begin{equation}
\theta=\frac12+\frac{u}{2\sqrt v},\qquad u=O(1).
\end{equation}
Then Eq.~\eqref{eq:sm-heat} gives, uniformly for $u$ in compact subsets of
$\mathbb R$,
\begin{equation}
\frac{e^{v/2}\,\E_{y_0}[S_A^\theta]}{\sqrt{S_0}}
=\mathcal F(u)+o(1),\qquad
\mathcal F(u)=e^{u^2/2}\operatorname{erfc}\!\left(\frac{u}{\sqrt2}\right).
\label{eq:sm-critical-scaling}
\end{equation}
Indeed, the left-hand side before taking the limit is
\begin{equation}
2^{-u/(2\sqrt v)}
\E\!\left[\cosh(y_0+\sqrt v\,Z)^{-u/\sqrt v}\right],
\qquad Z\sim N(0,1),
\end{equation}
and $v^{-1/2}\ln\cosh(y_0+\sqrt v Z)\to|Z|$.  Hence the limit is
$\E[e^{-u|Z|}]=e^{u^2/2}\operatorname{erfc}(u/\sqrt2)$.  The identities
$\mathcal F(0)=1$,
$\mathcal F(u)\sim\sqrt{2/\pi}/u$ as $u\to+\infty$, and
$\mathcal F(u)\sim2e^{u^2/2}$ as $u\to-\infty$ match, respectively, the
exact half-moment, the integrable spectral edge, and the two moving Gaussian
saddles.  This scaling resolves the QND moment-order layer; it is not a claim
of finite-action optimality over all switching policies.

\begin{figure}[t]
\centering
\includegraphics[width=0.92\textwidth]{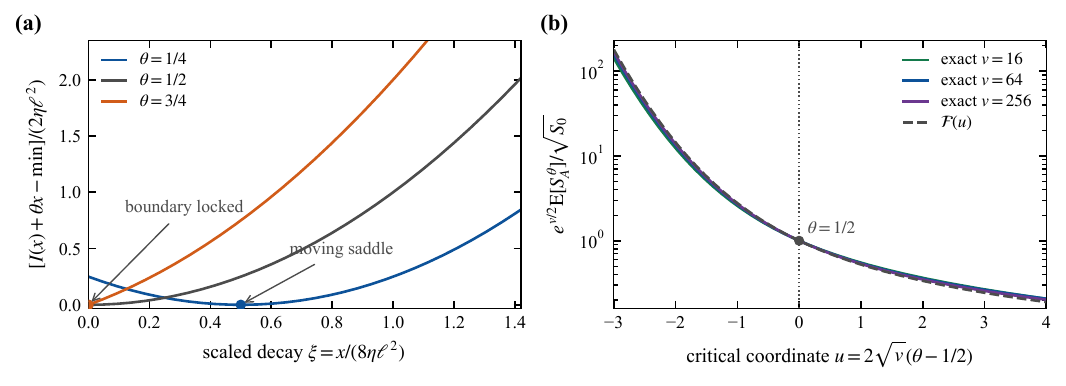}
\caption{Full fixed-QND anatomy of the half-moment transition.
(a) The tilted large-deviation potential has minimizer
$\xi_\theta=\max\{1-2\theta,0\}$: the moving saddle reaches the physical boundary
at $\theta=1/2$ and remains locked there.
(b) Finite-action moments at $S_0=0.455$ collapse onto the scaling function
$\mathcal F(u)$ in Eq.~\eqref{eq:sm-critical-scaling}.  Curves are converged
numerical quadrature of the exact QND kernel; no trajectory time discretization
is used.}
\label{fig:sm-critical}
\end{figure}

\emph{Late QND--unbiased crossing.---}
The efficiency--action crossover discussed in the Letter is distinct from the
moment-order layer above.  For $\theta=1$, set $\delta=1-\eta$ and
$y_0=\operatorname{arctanh}\sqrt{1-2S_0}$.  Expanding the Gaussian in
Eq.~\eqref{eq:sm-heat} and using
$\int_{\mathbb R}\operatorname{sech}y\,dy=\pi$ and
$\int_{\mathbb R}y^2\operatorname{sech}y\,dy=\pi^3/4$ gives
\begin{equation}
\E[S_A]_{\rm QND}
=\frac{\sqrt{\pi S_0}}{2\sqrt v}\,e^{-v/2}
\left[1-\frac{B}{v}+O(v^{-2})\right],
\qquad
B=\frac12\left(y_0^2+\frac{\pi^2}{4}\right).
\label{eq:sm-qnd-tail-refined}
\end{equation}
The always-unbiased solution is
\begin{equation}
S_A^{\rm ub}=\frac{\delta}{2}
+\left(S_0-\frac{\delta}{2}\right)e^{-v/\eta}.
\label{eq:sm-unbiased-v}
\end{equation}
At the nonzero late crossing as $\delta\downarrow0$, the transient in
Eq.~\eqref{eq:sm-unbiased-v} is negligible relative to the floor $\delta/2$.
With
\begin{equation}
W_\delta=W_0\!\left(\frac{\pi S_0}{\delta^2}\right),
\end{equation}
where $W_0$ is the principal Lambert function, the crossing therefore obeys
\begin{align}
v_\times
&=W_\delta-\frac{2B}{1+W_\delta}+O(W_\delta^{-2}),
\label{eq:sm-vcross}\\
A_\times
&=\frac{v_\times}{4\eta\ell^2}\nonumber\\
&=\frac{2L-\ln(2L)+\ln(\pi S_0)+O(\ln L/L)}
{4\eta\ell^2},
\qquad L=\ln\frac1\delta .
\label{eq:sm-across}
\end{align}
Thus the leading logarithm has a negative $\ln\ln(1/\delta)$ correction.
Equations~\eqref{eq:sm-vcross}--\eqref{eq:sm-across} describe the late
intersection of the always-QND and always-unbiased protocols.  They do not
identify a finite-horizon optimum over protocols that switch between the two.

\begin{figure}[t]
\centering
\includegraphics[width=0.92\textwidth]{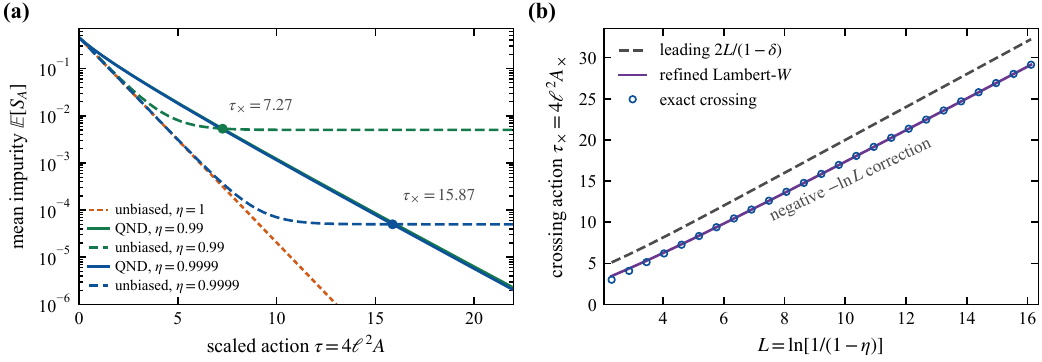}
\caption{Finite-action resolution of the singular efficiency limit, for
$S_0=0.455$ and scaled action $\tau=4\ell^2A$.
(a) Exact QND mean impurity (solid) and always-unbiased flow (dashed); circles
mark their nonzero late crossings.
(b) Exact crossing actions (circles) follow the refined Lambert-$W$ expansion
of Eqs.~\eqref{eq:sm-vcross}--\eqref{eq:sm-across} (purple), including the
negative $\ln L$ correction, rather than its slowly convergent leading term
(gray).  The comparison concerns the two fixed endpoint protocols.}
\label{fig:sm-crossover}
\end{figure}

Figure~\ref{fig:sm-crossover} shows the two finite-action laws and their late
crossings directly: an always-unbiased protocol initially shadows the ideal
curve before stalling, and the exact crossing actions follow the refined
Lambert-$W$ expansion of Eqs.~\eqref{eq:sm-vcross}--\eqref{eq:sm-across},
including the negative $\ln L$ correction.

\section{First-passage bound and exact QND hitting time}
\label{sm:fp}

Fix a qubit with $\ell>0$, $0<\eta\le1$, and $0<\epsilon<S_0$;
the dimension-independent extension is given at the end of this section.  Let
$T_\epsilon=\inf\{t:S_t\le\epsilon\}$ and adopt the cost convention
\begin{equation}
\mathcal C_\epsilon=
\begin{cases}A(T_\epsilon),&T_\epsilon<\infty,\\ +\infty,&T_\epsilon=\infty.
\end{cases}
\label{eq:sm-fpcost}
\end{equation}
Thus a protocol that has a nonzero probability of never reaching the target has
infinite expected cost.  It suffices below to consider successful protocols with
$T_\epsilon<\infty$ almost surely and $\E A(T_\epsilon)<\infty$; otherwise the
bound is immediate.

This corollary rests on a supermartingale built from $\ln S$, which we construct in full here.
It\^o's formula for $\ln S$ gives
\begin{equation}
-\mathcal{A}\ln S = M\Big[\,\frac{2g(x,S)}{S} + 8\eta\, x\beta\,\Big],
\label{eq:sm-lngen}
\end{equation}
the $\theta\to0$ member of the family (\ref{eq:sm-momgen}) after division by $\theta$,
in which the It\^o correction enters with full weight: randomness assists the logarithmic descent. The
bracket is affine in $x$; the aligned-minus-unbiased difference factorizes as
\begin{equation}
\big[\cdot\big]_{x=\ell^{2}}-\big[\cdot\big]_{x=0}
= 2\ell^{2}\,(1-2S)\Big[(3\eta-1)+(1-\eta)\frac{1+S}{S}\Big]
\;\ge\; 4\ell^{2}(1-2S)\;\ge\;0,
\end{equation}
using $(1+S)/S\ge3$ on $S\le\tfrac12$, so the aligned basis maximizes the descent rate
for every $(S,\eta)$. The aligned value,
$2\ell^{2}[(1+\eta)+(3\eta-1)\beta+2(\eta-1)S]
=8\eta\ell^2(1-S)$, is decreasing in $S$, with supremum
$8\eta\ell^{2}$ at $S\to0$. Hence, for every adapted protocol,
$\ln(1/S_t) - 8\eta\ell^{2}A(t)$ is a supermartingale. On the stopped interval
$[0,T_\epsilon]$ the process $\ln(1/S)$ is bounded
($\epsilon\le S\le\tfrac12$).  Stop first at
\begin{equation}
\tau_n=T_\epsilon\wedge n\wedge\inf\{t:A(t)\ge n\}.
\end{equation}
The martingale part is square integrable up to $\tau_n$ because its quadratic
variation is bounded by a constant times $A(\tau_n)\le n$.  Optional sampling
therefore applies.  As $n\to\infty$, continuity of the diffusion gives
$S_{T_\epsilon}=\epsilon$; dominated convergence applies to the bounded logarithm,
whereas monotone convergence applies to $A(\tau_n)\uparrow A(T_\epsilon)$.  Hence
\begin{equation}
8\eta\ell^{2}\,\E\big[A(T_\epsilon)\big]\;\ge\;\E\big[\ln(1/S_{T_\epsilon})\big]-\ln(1/S_0)
\;=\;\ln(S_0/\epsilon),
\end{equation}
which is an additional first-passage corollary beyond the Letter's main line.

The aligned protocol also permits an exact finite-threshold calculation.  In the
action clock its generator is
\begin{equation}
\mathcal L_A=\frac{\kappa\eta}{2}\partial_y^2
 +\kappa\eta\tanh y\,\partial_y,
\qquad \kappa=4\ell^2.
\end{equation}
Put
\begin{equation}
Y=\operatorname{arccosh}\frac{1}{\sqrt{2\epsilon}},
\qquad
y_0=\operatorname{arctanh}\sqrt{1-2S_0}.
\end{equation}
The target is the pair of boundaries $y=\pm Y$, and
$\mathcal L_A[y\tanh y]=\kappa\eta$.  Dynkin's formula for the stopped process
therefore gives the exact mean action
\begin{equation}
\displaystyle
\E A_{\rm QND}(T_\epsilon)
=\frac{Y\tanh Y-y_0\tanh y_0}{4\eta\ell^2}.
\label{eq:sm-exactfp}
\end{equation}
Equation~(\ref{eq:sm-exactfp}) is the exact QND mean hitting-time formula of
Li \emph{et al.}~\cite{Li2013}, rewritten in our action convention.  In the absence
of extrinsic decoherence, that work also proved QND measurement globally optimal for
the qubit first-passage objective at every $0<\eta\leq1$.  The additional statement
below is the dimension-independent speed bound, not a new qubit first-passage optimum.
In particular,
$\E A_{\rm QND}(T_\epsilon)=\ln(S_0/\epsilon)/(8\eta\ell^2)+O(1)$ for fixed
$S_0$ and $\epsilon\downarrow0$, so the universal bound is saturated at leading
logarithmic order, not as a finite-$\epsilon$ equality.

At $\eta=1$ the unbiased protocol instead obeys
$dS/dA=-4\ell^2S$ and therefore has the exact cost
$A_{\rm ub}=\ln(S_0/\epsilon)/(4\ell^2)$.  Combining this expression with
Eq.~(\ref{eq:sm-exactfp}) shows
\begin{equation}
\frac{A_{\rm ub}}{\E A_{\rm QND}}\longrightarrow2
\qquad(\epsilon\downarrow0),
\end{equation}
but the ratio is not exactly two at a finite threshold.  This unit-efficiency
first-passage reversal was identified in Ref.~\cite{WisemanRalph2006}.

\emph{All dimensions.---} The first-passage bound extends to every finite
dimension as the $\theta\to0$ member of the general-$d$ moment envelope. By the
reduction theorem (Sec.~\ref{sm:ddimmoment}), the pointwise generator bound
$-\mathcal{A}S^{\theta}/S^{\theta}\le 8\theta(1-\theta)\,\eta\ell^{2}M$ for
$\theta\le\tfrac12$ holds in all $d$.
Dividing by $\theta$ and letting $\theta\to0$ (the It\^o correction surviving with full
weight, $8(1-\theta)\to8$) gives
\begin{equation}
-\mathcal{A}\ln S \;\le\; 8\eta\ell^{2}M \qquad\text{in every dimension},
\end{equation}
so $\ln(1/S_t)-8\eta\ell^{2}A(t)$ is a supermartingale for every adapted protocol at
any $d$. Use the same infinite-cost convention (\ref{eq:sm-fpcost}).  For a
successful protocol of finite expected action, stop at $\tau_n$ as above.
The bound $\epsilon\le S_t\le1-1/d<1$ before the stop permits dominated
convergence for $\ln(1/S_{\tau_n})$, while monotone convergence applies to
$A(\tau_n)$.  The argument therefore gives, for $0<\eta\le1$,
\begin{equation}
8\eta\ell^{2}\,\E\big[A(T_\epsilon)\big]\;\ge\;\ln(S_0/\epsilon),
\qquad\text{i.e.}\qquad
\E\big[A(T_\epsilon)\big]\;\ge\;\frac{\ln(S_0/\epsilon)}{8\eta\ell^{2}}
\quad\text{for all }d,
\end{equation}
an additional first-passage corollary, valid at every finite dimension.

\section{Qubit fixed-threshold certificates and moment-ceiling attainment}
\label{sm:frozen}

For a qubit and every $0<\eta<1$, we prove a family of fixed-threshold
certificates.  The scalar case $\ell=0$ is trivial, so assume $\ell>0$.  At each
fixed $\omega>0$ the threshold $s_Y$ below is fixed and its
protocol-uniform survival certificate has rate
$\lambda_\omega=2\eta\ell^2(1+\omega^2)$.  Taking the long-action limit first and
only then $\omega\downarrow0$ recovers the moment ceiling of
Sec.~\ref{sm:alldfreeze}.  Because QND attains that ceiling from every mixed qubit
initial state when $\theta\geq\tfrac12$, this also completes the exact qubit
moment-optimality result.  We do not claim a fixed-threshold survival exponent
equal to $2\eta\ell^2$.

We use the action clock $a=A(t)$, so the displayed generator below is per unit
action, and work in the coordinate
$y=\operatorname{arctanh}\sqrt{\beta}$ of Sec.~\ref{sm:exact}. For an arbitrary
adapted basis protocol $x_t\in[0,\ell^{2}]$, It\^o's formula applied to $y(S)$ with
the impurity drift and diffusion of Sec.~\ref{sm:param} gives, in the interior
$y>0$, the controlled diffusion
\begin{equation}
dy \;=\; \frac{g(x,S)-2\eta x S(1-3\beta)}{S\sqrt{\beta}}\;da
\;+\; 2\sqrt{\eta x}\,dW_a,
\label{eq:sm-ydrift}
\end{equation}
whose drift and squared diffusion are both \emph{affine in $x$}, the same structural
fact used throughout.  At $y=0$ the radial coordinate is reflected (equivalently,
one may use a signed lift), but the resulting local-time term vanishes for the even
verification function below because $\psi'(0)=0$.  Thus the displayed generator
applies at the boundary by continuity.  Fix $\omega\in(0,\sqrt{1-\eta}\,]$, set $Y=\pi/(2\omega)$,
$s_{Y}=\tfrac12\operatorname{sech}^{2}Y$, and define on $S\in[s_{Y},\tfrac12]$ the
verification function and rate
\begin{equation}
\psi(S) \;:=\; \operatorname{sech}(y)\cos(\omega y)\big|_{y=y(S)},
\qquad
\lambda \;:=\; 2\eta\ell^{2}\,(1+\omega^{2}) .
\label{eq:sm-psi}
\end{equation}
$\psi$ is an even, hence smooth, function of $y$, with $0\le\psi\le1$ and
$\psi(s_{Y})=0$. The core of the proof is the protocol-uniform generator inequality
\begin{equation}
\mathcal{A}_{x}\,\psi \;\ge\; -\lambda\,\psi
\qquad \text{for every } x\in[0,\ell^{2}],\; S\in[s_{Y},\tfrac12] .
\label{eq:sm-genineq}
\end{equation}
By affinity in $x$ it suffices to verify the two endpoints. At the aligned endpoint
$x=\ell^{2}$, the ground-state transform of Sec.~\ref{sm:exact} gives the exact
conjugation $\mathcal{A}_{\ell^{2}}[\operatorname{sech}(y)\,v(y)] =
2\eta\ell^{2}\operatorname{sech}(y)\,[v''(y)-v(y)]$; with $v=\cos(\omega y)$ this is
the eigenrelation $\mathcal{A}_{\ell^{2}}\psi=-\lambda\psi$, so
Eq.~(\ref{eq:sm-genineq}) holds there with equality. At the unbiased endpoint $x=0$
the motion is deterministic with drift
$D_{0}(y)=2\ell^{2}\,[1-(1-\eta)\cosh^{2}y]/\tanh y
=2\ell^2(\eta-u)/\tanh y$, where
$u=(1-\eta)\sinh^2y$.  Since
\begin{equation}
\sqrt{1-\eta}\,\operatorname{arctanh}\sqrt\eta<1<\frac{\pi}{2},
\end{equation}
(use $\operatorname{arctanh}r\le r/\sqrt{1-r^2}$), the choice
$\omega^2\le1-\eta$ puts the entire region $D_0>0$ inside $0\le\omega y<\pi/2$.
If $D_0\le0$, then $\psi'(y)\le0$ and
$\mathcal A_0\psi=D_0\psi'\ge0\ge-\lambda\psi$.  It remains to treat
$D_0>0$, equivalently $0\le u<\eta$.  After substituting $D_0$ and
$\psi'=-\operatorname{sech}y[\tanh y\cos(\omega y)+
\omega\sin(\omega y)]$, Eq.~(\ref{eq:sm-genineq}) is equivalent to
\begin{equation}
\big(u+\eta\omega^{2}\big)\cos(\omega y) \;\ge\; (\eta-u)\,\omega\sin(\omega y)\coth y,
\label{eq:sm-star}
\end{equation}
including its continuous limit at $y=0$.  Here is the complete closure of this
inequality.  Put $w=\omega^2$.  The elementary bounds
$\sin(\omega y)\le\omega y$,
$\coth y\le y^{-1}+y/3$, and
$\cos(\omega y)\ge1-wy^2/2$ give, for the left side minus the right side of
Eq.~(\ref{eq:sm-star}),
\begin{align}
\Delta
&\ge (u+\eta w)\Big(1-\frac{wy^2}{2}\Big)
 -(\eta-u)w\Big(1+\frac{y^2}{3}\Big)\notag\\
&=u(1+w)-\frac{wy^2}{6}\,[u+2\eta+3\eta w].
\label{eq:sm-star-chain}
\end{align}
On this region $0\le u\le\eta$, so
$u+2\eta+3\eta w\le3\eta(1+w)$.  Moreover,
$u=(1-\eta)\sinh^2y\ge(1-\eta)y^2$.  Therefore
\begin{equation}
\Delta\ge y^2(1+w)\Big[(1-\eta)-\frac{\eta w}{2}\Big]\ge0,
\end{equation}
where the final inequality uses $w\le1-\eta$.  This proves
Eq.~(\ref{eq:sm-genineq}) at the unbiased endpoint and hence, by affinity, at
every $x\in[0,\ell^2]$.

\begin{figure}[t]
\centering
\includegraphics[width=0.92\textwidth]{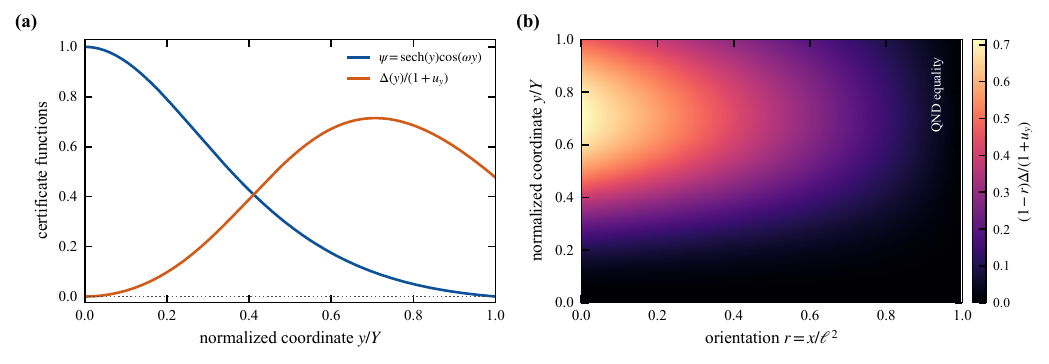}
\caption{Analytic fixed-threshold certificate at $\eta=3/4$,
$\omega=1/2$, and $Y=\pi$.
(a) The nonnegative verification function $\psi$ vanishes at the stopping
threshold, while the normalized endpoint residual
$\Delta/(1+u_y)$ remains nonnegative.
(b) The normalized all-orientation residual
$(1-r)\Delta/(1+u_y)$ is nonnegative throughout the strip.  Its exact vanishing
on the QND edge $r=1$ displays the eigenrelation used in the proof; affinity in
$r$ then certifies every predictable intermediate orientation.}
\label{fig:sm-survival}
\end{figure}

For transparency, Fig.~\ref{fig:sm-survival} plots the exact verification function
and the generator residual for one admissible certificate.  With
$r=x/\ell^2$ and $u_y=(1-\eta)\sinh^2y$, the identity behind the plot is
\begin{equation}
\frac{\mathcal A_x\psi+\lambda\psi}
{2\ell^2\operatorname{sech}y}
=(1-r)\Delta_{\eta,\omega}(y),
\label{eq:sm-residual-factorization}
\end{equation}
where $\Delta_{\eta,\omega}$ is the left side minus the right side of
Eq.~(\ref{eq:sm-star}).

\begin{figure}[!t]
\centering
\includegraphics[width=0.92\textwidth]{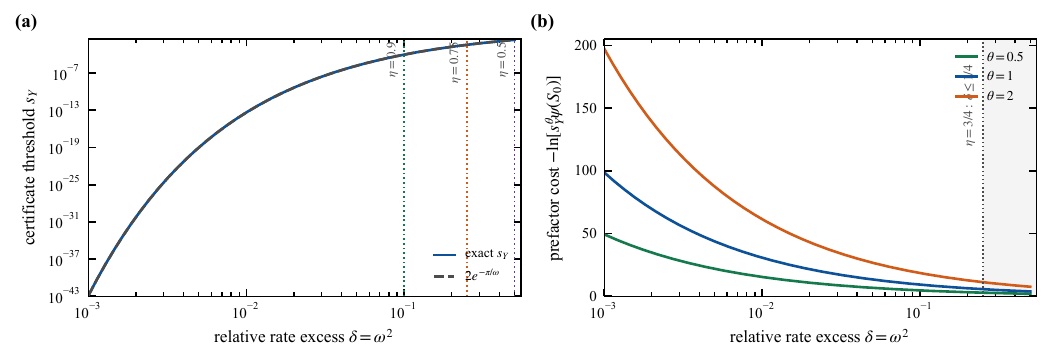}
\caption{Threshold and prefactor cost of approaching the frozen rate.
(a) The exact threshold in Eq.~(\ref{eq:sm-threshold-tradeoff}) and its
small-$\delta$ asymptote; vertical lines mark the admissible endpoints
$\delta=1-\eta$.
(b) For $S_0=0.455$, the prefactor cost
$-\ln[s_Y^\theta\psi_\omega(S_0)]$ diverges as $\delta\downarrow0$.
The $\eta=3/4$ marker corresponds to the certificate in
Fig.~\ref{fig:sm-survival}.  Hence every fixed threshold in this certificate family
($\omega^2\le1-\eta$) has
$\lambda_\omega>2\eta\ell^2$, and the sharp ceiling is recovered only by taking
$\bar A\to\infty$ before $\omega\downarrow0$.  These are rigorous certificate
rates in lower bounds, not claimed exact finite-threshold survival exponents or
finite-horizon optima.}
\label{fig:sm-threshold-tradeoff}
\end{figure}

To extract the survival bound, choose $\omega$ small enough that $s_Y<S_0$ and let
$\tau_Y=\inf\{t:S_t\le s_Y\}$.  The stopped process
$e^{\lambda A_{t\wedge\tau_Y}}\psi(S_{t\wedge\tau_Y})$ is a nonnegative local
submartingale.  Localization followed by bounded convergence (all factors are
bounded under the action cap below) gives, for every adapted protocol satisfying
$A_T\le\bar A$ almost surely,
\begin{equation}
\psi(S_0)
\le \E\!\left[e^{\lambda A_T}\psi(S_T)\mathbf1_{\{\tau_Y>T\}}\right]
\le e^{\lambda\bar A}\Pr(\tau_Y>T).
\label{eq:sm-survival}
\end{equation}
On $\{\tau_Y>T\}$ one has $S_T\ge s_Y$.  Thus, for every $\theta>0$,
\begin{equation}
\E\big[S_{T}^{\,\theta}\big] \;\ge\; s_{Y}^{\,\theta}\;\psi(S_{0})\;
e^{-2\eta\ell^{2}(1+\omega^{2})\,\bar A} .
\label{eq:sm-frozenbound}
\end{equation}
The cost of approaching the sharp rate can be read off explicitly.  Put
\begin{equation}
\delta_\omega:=\frac{\lambda_\omega}{2\eta\ell^2}-1=\omega^2,
\qquad
s_Y=\frac12\operatorname{sech}^2\!\left(\frac{\pi}{2\sqrt{\delta_\omega}}\right),
\qquad 0<\delta_\omega\leq1-\eta.
\label{eq:sm-threshold-tradeoff}
\end{equation}
As $\delta_\omega\downarrow0$,
$s_Y\sim2e^{-\pi/\sqrt{\delta_\omega}}$.  Thus the prefactor
$C_{\theta,\omega}:=s_Y^\theta\psi_\omega(S_0)$ obeys
$-\ln C_{\theta,\omega}=\pi\theta/\sqrt{\delta_\omega}+O(1)$ for every fixed
mixed $S_0$.  Figure~\ref{fig:sm-threshold-tradeoff} displays this
nonperturbative rate--prefactor tradeoff for the whole certificate family.

For each \emph{fixed} $\omega>0$, the prefactor
$s_Y^\theta\psi(S_0)$ is positive and independent of $\bar A$.  We therefore
first take $\bar A\to\infty$ in Eq.~(\ref{eq:sm-frozenbound}), obtaining an
upper bound $2\eta\ell^2(1+\omega^2)$ on any achievable decay exponent, and only
then take $\omega\downarrow0$.  (Reversing these limits would be invalid because
$s_Y\to0$.)  No adaptive qubit protocol can consequently achieve an exponent
larger than $2\eta\ell^{2}$.  Since plain measurement attains this value
for every $\theta\ge\tfrac12$ (Sec.~\ref{sm:exact}), the frozen asymptotic
ceiling is optimal on the entire range $\theta\ge\tfrac12$ at every $0<\eta<1$.  For each fixed
$\omega$, Eq.~(\ref{eq:sm-survival}) is an event-level certificate at the
threshold-specific rate $\lambda_\omega>2\eta\ell^2$; the value
$2\eta\ell^2$ appears only after the ordered limits above.  The all-dimensional
result of Sec.~\ref{sm:alldfreeze} gives the same moment-exponent ceiling, but it
does not give a fixed-threshold certificate; Sec.~\ref{sm:detlaw} separately shows
that generic full-rank $d>2$ states do not attain that uniform constant near the
pivot.  Two consistency checks: at
$\eta=1$ the admissible window $\omega\le\sqrt{1-\eta}$ collapses, precisely where the
unbiased basis escapes the freeze and attains $4\theta\ell^{2}$; and
Eq.~(\ref{eq:sm-survival}) itself is the corresponding qubit
threshold-and-rate survival certificate.

\section{Numerical verification}
\label{sm:numerics}

The analytic statements of the Letter and this Supplement were checked
numerically; all checks pass.  These checks illustrate the theorems and are independent
of the proofs, which are analytic.

The result figures are generated from the closed-form rate curves and exact
Gaussian quadrature of the QND heat kernel, so they carry no discretization
error.  For the canonical qutrit spectrum $(-\ell,0,\ell)$ at $\eta=0.9$ the
state-space-uniform ceiling $\Lambda_{\rm unif}(\theta)/\ell^2=8\eta\theta(1-\theta)$
for $0<\theta\le\tfrac12$ and $2\eta$ for $\theta\ge\tfrac12$, attained on embedded
extremal rank-two QND faces, and the full-rank determinant ceiling
$\Lambda_{\rm det}(\theta)/\ell^2=\tfrac{4\eta}{3}\max(1,\tfrac{3\theta}{2})$ (from
$V_L=\tfrac23\ell^2$) are strictly separated on
$(1-1/\sqrt3)/2<\theta<1$, with the determinant kink at $\theta=\tfrac23$ and
equality at both endpoints; for $L_d=\operatorname{diag}(-\ell,0^{d-2},+\ell)$ the
strict-obstruction interval $(1-\sqrt{1-2/d})/2<\theta<(d-1)/2$ collapses to the
half-moment pivot at $d=2$ and is nonempty for every $d>2$.

Across random states and observables for $2\le d\le8$ and $0.02\le\theta\le4$, the
quartic inequality, the master inequality $C_{\sqrt S}$, and the moment-generator
bound hold, and their equality cases occur exactly on the rank-two supports of the
classification.  A near-pure scan to $d=20$ and $S\sim10^{-10}$ confirms the
rigidity bound $T_2\le12\ell^2S^2+8S\Delta$ and the local drift bound
$B\le24\eta\ell^2S^2$, and the deterministic zero-innovation qutrit direction
realizes the $O(S^2)$ drift that makes the scalar fixed-threshold argument
qubit-specific.  The matrix-It\^o determinant generator matches
$-2\eta V_L+(1-\eta)K/d$ with $K\ge0$, and the canonical qutrit value
$V_L=\tfrac23\ell^2$ is reproduced.

\begin{figure}[t]
\centering
\includegraphics[width=0.92\textwidth]{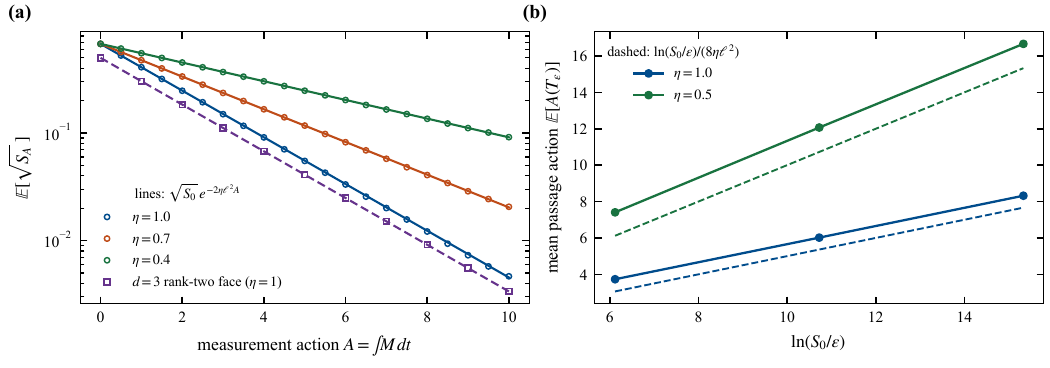}
\caption{Independent trajectory checks of two bounds.
(a) For a qubit, Monte Carlo estimates (circles) of $\E[\sqrt{S_A}]$ follow the
exact QND law (solid lines) at three efficiencies.  The purple squares are analytic
samples of a $d=3$ realization with zero population in the intermediate eigenspace;
they realize the rank-two embedded sharpness construction, which saturates the uniform
ceiling only on that rank-two face; a generic full-rank qutrit stays strictly below it.  The qubit curves start
at $S_0=0.455$ and the embedded $d=3$ samples at $S_0=\tfrac14$.
(b) QND mean first-passage actions (points) approach the all-dimensional bound
(dashed) from above and match its predicted leading-logarithmic slope.}
\label{fig:sm-verification}
\end{figure}

Figure~\ref{fig:sm-verification} shows an independent trajectory check.  Monte
Carlo estimates of $\E[\sqrt{S_A}]$ follow the exact QND law at three efficiencies,
and the mean first-passage actions approach the dimension-independent bound
$\ln(S_0/\epsilon)/(8\eta\ell^2)$: their fitted leading-logarithmic slopes are
$0.50$ at $\eta=1$ and $1.01$ at $\eta=\tfrac12$, matching the predicted $0.5$ and
$1$ at $\ell^2=\tfrac14$.

\end{document}